\documentclass[preprint, amsmath,amssymb, aps,prd,showpacs]{revtex4-1}

\usepackage{graphicx, graphics, color}
\usepackage[T1]{fontenc} 
\usepackage{amsmath, amssymb, bm,mathrsfs}
\usepackage{epsfig}
\usepackage{dcolumn}
\usepackage{bm}
\usepackage{tikz}
\usepackage{dcolumn}
\usepackage[mathlines]{lineno}
\usepackage{subfig}
\usepackage{float}
\usepackage[colorlinks=true,citecolor=blue]{hyperref}
\usepackage[normalem]{ulem}

\newcommand{\be}{\begin{equation}}
\newcommand{\ee}{\end{equation}}
\newcommand{\ben}{\begin{eqnarray}}
\newcommand{\een}{\end{eqnarray}}

\newcommand{\cO}{{\cal O}}

\newcommand{\p}{\partial}
\newcommand{\na}{\nabla}

\newcommand{\ep}{\epsilon}

\newcommand{\ga}{\gamma}

\newcommand{\talpha}{{\tilde \alpha}}
\newcommand{\tbeta}{{\tilde \beta}}

\newcommand{\zz}{\mathbb{Z}_2}

\graphicspath{{figs/}}


\begin{document} 
\title{Anomalous
Hall conductivity of the holographic $\mathbb{Z}_2$ Dirac semimetals}

\author{Bartlomiej Kiczek} 
\email{bkiczek@kft.umcs.lublin.pl}
\author{Marek Rogatko} 
\email{rogat@kft.umcs.lublin.pl}
\author{Karol I. Wysokinski}
\email{karol@tytan.umcs.lublin.pl}
\affiliation{Institute of Physics, 
Maria Curie-Sklodowska University, 
pl.~Marii Curie-Sklodowskiej 1,  20-031 Lublin,  Poland}

\date{\today}

\begin{abstract}
 The anomalous spin Hall conductivity in the holographic model of Dirac semimetals with two Dirac nodes protected by the crystal symmetry has been elaborated. Such system besides the chiral anomaly possesses another anomaly which is related to the $\mathbb{Z}_2$ topological charge of the system. 
The holographic model of the system contains matter action with two $U(1)$-gauge fields as well as the appropriate combination of the Chern-Simons gauge terms. 
 We also allow for the coupling of two gauge fields  {\it via} the kinetic mixing parametrised by the  coupling $\alpha$. The holographic approach in the probe limit enables us to obtain Hall conductivity. The aim of this work is to describe the phase transitions in the $\mathbb{Z}_2$ Dirac semimetals between the topologically trivial and non-trivial phases. Interestingly the anomalous Hall conductivity plays a role of the order parameter of this phase transition. The holographically found prefactor of the Hall conductivity in the topologically non-trivial phase, depends on the coupling $\alpha$ and  the Chern - Simons couplings. 
\end{abstract}

\maketitle
\flushbottom



\section{Introduction}
\label{sec:intro}
Recently it has been argued that near the charge neutrality point in two and three dimensional systems with Dirac massless energy spectrum, a strong interacting plasma (or Dirac fluid) forms. The evidence of such kind of strong interacting charged fluid was observed in experiments indicating the violation of the Wiedemann-Franz  (WF) law \cite{cro16} or the appearance of the viscous flow of charge carriers \cite{sul19} in the extremely clean graphene near the charge neutrality point. On the theoretical side these facts initiated the emergence of the holographic generalisation of hydrodynamical approach to the aforementioned systems \cite{har07,luc16}. In order to quantitatively explain the experimental data on the thermal conductivity of graphene, the holographic model of strongly coupled plasma with two non-interacting $U(1)$-gauge currents was proposed \cite{seo17}. Two fields bounded, respectively, with the electron and hole currents in graphene with Fermi energy  coinciding with the Dirac point have been introduced. They allowed for the nearly perfect agreement with experiments.

The model  has been further generalised \cite{rog18} by allowing for the interaction of two $U(1)$-gauge currents via so-called {\it kinetic mixing term} with non-zero coupling constant between two fields. The thermoelectric  and magneto-transport properties of graphene have been studied. The Hall effect has been found in the standard geometry with the magnetic field perpendicular to the graphene plane,  with the electric field and temperature gradients lying in the plane but being perpendicular to each other. The DC-transport coefficients were calculated by the introduction of the axionic field, which provides momentum relaxation mechanism related to the finite mobility of  carriers. The auxiliary $U(1)$-gauge field played an important role and affected the kinetic and transport coefficients {\it via} the parameter $\alpha$, connected with the {\it kinetic mixing term}.

The holographic model with two interacting gauge currents predicts \cite{rog18} that the increase of $\alpha$-coupling constant leads to the increase of the width of normalised thermal conductivity with the parameter which bounds both $U(1)$ charges. Moreover, the dependence of the WF ratio on the $\alpha$-coupling constant  results in the changes of the width of curves and their heights. This dependence was found to be valid for all charge densities. The dependence of the Seebeck coefficient on the charge concentration, for the different values of mobilities was also studied and  a very good agreement with the experimental data was achieved.

After the experimental discovery of the three dimensional analogs of graphene, the so called Weyl and Dirac semimetals \cite{armitage2018} it became clear that besides the relativistic massless spectrum the carriers in these systems possess quantum anomalies. In the case of Weyl semimetals it is the chiral anomaly which at the quantum level shows up as a non-conservation of the chiral charge. It has been also revealed that the class of Dirac semimetals with two Dirac nodes at the Fermi level separated in momentum space along the crystallographic rotation axis besides the chiral charge possesses another topological charge and the corresponding anomaly. The underlying symmetry and anomaly, after \cite{burkov2016} we call the $\mathbb{Z}_2$ anomaly and  the corresponding  systems the $\mathbb{Z}_2$ Dirac semimetals. These anomalies are related to chirality  and spin degrees of freedom of charge carriers in Dirac cones. The detailed physical discussion of these issues can be found in \cite{burkov2016}. While the positive longitudinal {\it charge} magnetoconductivity has been predicted as an experimental verification of the chiral anomaly in Weyl semimetals, the positive longitudinal {\it spin} magnetoconductivity is the proposed \cite{deng2019} smoking gun of the $\mathbb{Z}_2$ anomaly in the topological Dirac semimetals.

The extension of the aforementioned analysis \cite{seo17,rog18} of transport in graphene towards  three dimensional Dirac systems was provided in \cite{rog18a}. 
The hydrodynamical model of the Dirac $\mathbb{Z}_2$ semimetal  was studied in the theory 
with two interacting gauge fields. The implementation of chiral anomaly and $\zz$ topological charge \cite{rog18b} was taken into account, where the topological charge was described by the anomaly term in the additional gauge field sector. The existence of $\zz$ topological charge modifies equations and leads to the appearance of the new kinetic coefficients connected with vorticity and magnetic field of the auxiliary field. 

In \cite{rog19} the magneto-transport coefficients were found for the five-dimensional Chern-Simons generalisation of the presented holographic model.  The model in question enables one to describe the holographic Dirac semimetals with $\zz$ symmetry and leads to the positive longitudinal magnetoconductivity, at large \textbf{B} fields, with small region around \textbf{B}=\textbf{0} characterised by 
a negative magnetoconductivity, being in agreement with some experimental data. One also should remark that the model with two coupled vector fields, was used in a generalisation of 
p-wave superconductivity, for the holographic model of ferromagnetic superconductivity \cite{amo14}. 
 
The holographic model of Weyl semimetal which encoded the axial charge dissipation effect was elaborated in \cite{lan16,lan16a}. It has been revealed that varying the mass parameter the model underwent a sharp crossover at small temperature from a topologically non-trivial state to a trivial one. The holographic renormalisation group flow was a helpful device in the interpretation of the results, leading to the restoration of the time reversal symmetry at the end point of the renormalisation flow in the trivial phase.

\subsection{Motivation of the paper}
The main motivation standing behind our studies is to envisage how the quantum phase transition from topologically non-trivial to trivial phase looks like, for the Dirac $\zz$ semimetals. It will be the key point to visualise the role of the auxiliary $U(1)$-gauge field and especially the coupling between the two gauge fields in question, in the studied process.
Our holographic model possesses the Chern-Simons and {\it kinetic mixing} terms, binding various combinations of $U(1)$-gauge field strengths, with the adequate coupling constants.  The topologically non-trivial $\zz$ Dirac semimetal phase will be characterised by appearance of spontaneous Hall conductivity. The results can be interpreted in terms of the holographic renormalisation group flow.

 In order to obtain the Hall effect in topologically trivial system one needs the magnetic field or breaking the time reversal symmetry by other means. One of the proposals, presented in \cite{lan16},
 is to have a closer  look at the Weyl semimetal with two Weyl nodes of left and right-chirality, separated in the wave vector space, by a spatial vector $b_\mu$ (in the present convention $b_\mu b^\mu >0$). This separation causes the breaking of time reversal symmetry.
In the presence of massive fermions $M$ the parameter $b$ separating Weyl nodes along $z$ direction in momentum space gets modified to the effective value $b_{eff}$, which for the linear spectrum and for $b>M$  becomes 
$b_{eff}=\sqrt{b^2-M^2}$, while for $|b|<M$ the quantum phase transition to the gapfull state appears. The resulting gap is given by $\Delta= \sqrt{M^2-b^2}$. 
The holographic model of the described system is provided by the action \cite{lan16}
\ben \label{action1} \nonumber
S &=& \int dx^5 \sqrt{-g}\Bigg( R + \frac{12}{L^2} - \frac{1}{4}F_{\mu \nu} F^{\mu \nu} - \frac{1}{4}F^{(5)}_{\mu \nu} F^{(5)\mu \nu} \nonumber \\ 
 &+& \frac{\alpha}{3}\ep^{\mu \nu \rho \delta \tau} A^{(5)}_{\mu} (F^{(5)}_{\nu \rho}F^{(5)}_{\delta \tau}+3 F_{\nu \rho}~F_{\delta \tau}) - (D_\mu \Phi)^\dagger D^\mu \Phi - m^2 \Phi^\dagger \Phi
\Bigg),
\een
where the axial $U(1)$ symmetry is represented by the gauge field $A^{(5)}_\mu$ bounded to the field strength $F^{(5)}_{\mu \nu}$. This field is anomalous and it is the source of Chern-Simons
part of the above action.

\begin{figure}[h!]
\includegraphics[width=0.47\textwidth]{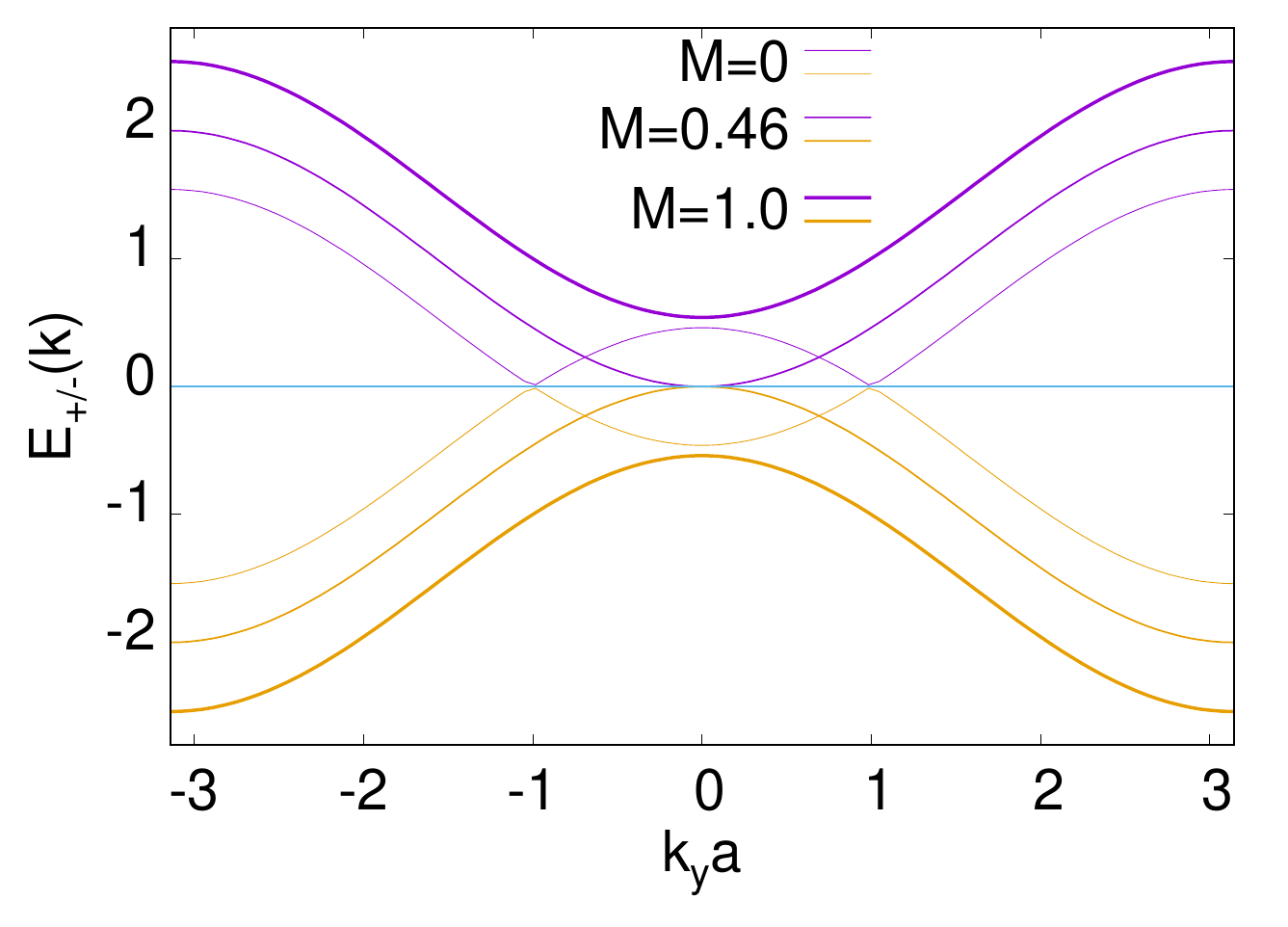}
\includegraphics[width=0.47\textwidth]{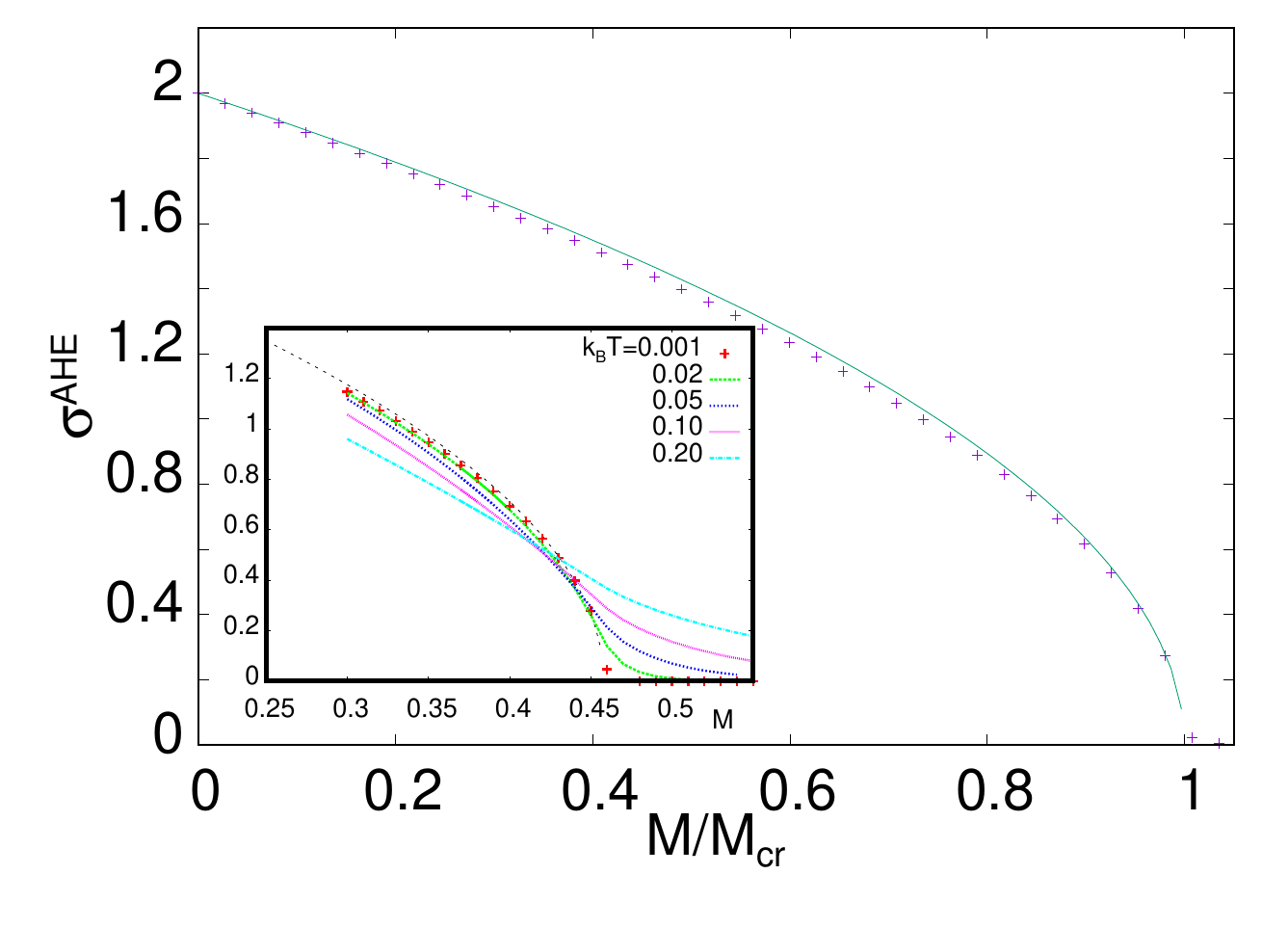}
\caption{Left panel shows the changes of the energy spectrum of the Weyl semimetal ploted for $k_x=k_z=0$ with increasing the fermionic mass $M$. In the right panel we show the dependence of the anomalous Hall effect on the mass parameter rescaled by its critical value for half-filed band. The full line represents analytical formula $2\sqrt{1-x}$. Note that $\sigma^{AHE}$ vanishes when the system develops the gap in the spectrum. The inset illustrates the detailed behaviour of $\sigma^{AHE}$ close to the critical value of $M$ for a number of temperatures.} 
\label{added}
\end{figure}

To better understand the above holographic action of the Weyl semimetal and its later generalisation  to the Dirac semimetal we shall present here some results on the anomalous Hall effect in a condensed matter lattice of, e.g., simple cubic symmetry (with lattice constant $a=1$)  described by the Weyl Hamiltonian with two nodes at the points $\mathbf{k}_W=(0,\pm b_y,0)$ 
\be
H(\mathbf{k})=d_x(\mathbf{k})\sigma_x+d_y(\mathbf{k})\sigma_y +d_z(\mathbf{k})\sigma_z,
\ee
with $d_x(\mathbf{k})=\gamma_x\sin(k_x)$, $d_y(\mathbf{k})=\gamma_y(\cos(k_y)-\cos(b_y))+\gamma_m(M+2-\cos(k_x)-\cos(k_z)$ and $d_z(\mathbf{k})=\gamma_z\sin(k_z)$. The parameter $M$ denotes the fermion mass \cite{bernevig2013}. The band structure along the $z$-axis for $k_x=k_z=0$, $\gamma_x=\gamma_y=\gamma_z=-\gamma_m=-1$ is shown in the left panel of figure \ref{added}. 
For $M=0$ one deals with massless fermions. Valence and conduction bands touch each other at Weyl points at $b_y=\pm 1$. For the critical value of the mass; $M_{cr}\approx 0.46$ the two bands touch each other at the single point at $k_y=0$, and effective distance vanishes $b_{eff}=0$. For larger $M$ the system develops real gap and becomes topologically trivial with vanishing Berry curvature and vanishing Hall conductivity. The right panel of the figure shows the anomalous Hall conductivity \cite{haldane2004} $\sigma^{AHE}=\sigma_{xz}= \frac{e^2}{\hbar}\frac{1}{N}\sum_{\lambda,\mathbf{k}}\Omega_{xz}^\lambda f_\lambda(\mathbf{k})$, where $\lambda$ enumerates the filled bands, $f_\lambda(\mathbf{k})$ is the equilibrium Fermi distribution function, $\Omega_{xz}^\lambda$ is the $xz$ component of the Berry curvature tensor of the band $\lambda$ and $N$ is the number of units cells in a crystal. The right panel of the Fig. \ref{added} shows the dependence of the Hall conductivity due to the Berry curvature on the mass term $M$. The conductivity vanishes for $M>M_{cr}$ signalling the transition from topologically non-trivial to trivial material. The conductivity of the half - filled band and for $M=0$ is given by the exact formula $\sigma^{AHE}=\frac{e^2}{\hbar}2b_y$ \cite{yang2011}.  The inset shows the detailed behaviour close to $M=M_{cr}$ for a number of temperatures.

The non-equilibrium effects in Dirac semimetals with $\zz$  topological charge \cite{burkov2016} require  a suitable generalisation of the above model, which has 
been done in \cite{rog18b,rog19}. Two different gauge fields were introduced, one of them coupled to charge of fermions and the other related to spin degrees of freedom. The low energy  Hamiltonian describing { inter alia} two materials argued to be Dirac semimetals and possessing  $\zz$ charge, namely Cd$_3$As$_2$ and Na$_3$Bi, is provided by the relation \cite{burkov2016} 
\be
H=a(\tau^x\sigma^zk_x-\tau^yk_y)+b(\textbf{k})\tau^z+\cO(\textbf{k}^3),
\label{lo}
\ee 
where $a$ and $b(\textbf{k})$ are in general $\textbf{k}$ dependent constants, $\cO(\textbf{k}^3)$ denotes terms of higher order in $\textbf{k}$, $\tau^i$ are Pauli matrices acting on orbital degrees of freedom, while $\sigma^i$ are Pauli matrices in the spin sector. It is seen that at low energy the $z$-component of spin is a good quantum number, as the operator $\sigma^z$ commutes with the low energy part of the Hamiltonian (\ref{lo}). 
 In Dirac $\zz$ semimetals the  separated cones differ by spin degree of freedom. 
On the holographic side the model requires suitable generalisations and this is a subject of the present paper.

\begin{figure}[h!]
\includegraphics[width=0.67\textwidth,angle=270]{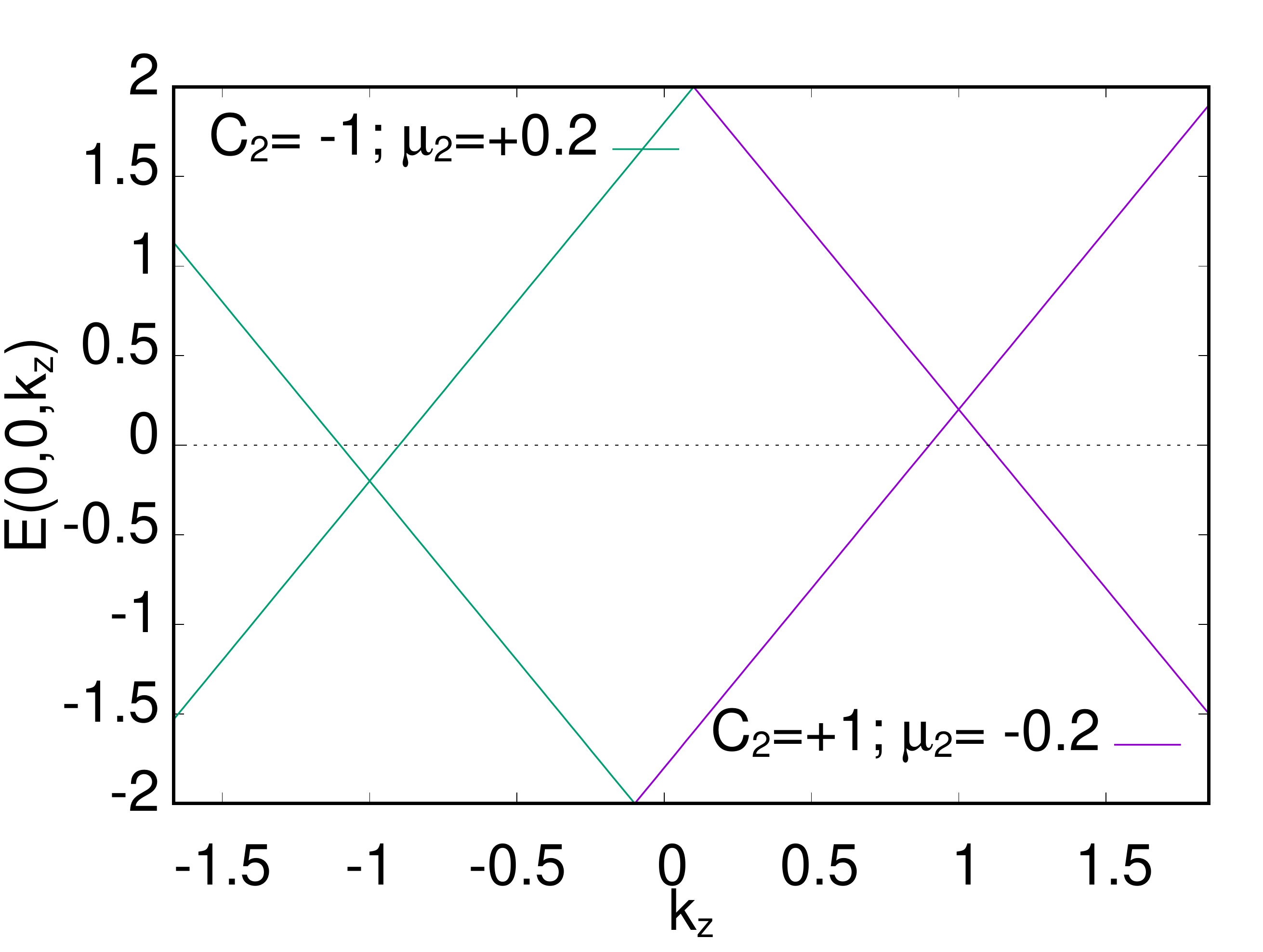}
\caption{The spectrum of electrons in Dirac semimetal with two separated  Dirac cones shifted in energy by the $\zz$ chemical potentials $\pm\mu_2$. Two Dirac nodes harbour different $\zz$ charges $C_2$ as indicated. As visible from Eq. (\ref{dsm-spin}) both spin directions lead to the same spectrum, which is thus spin degenerated.} 
\label{added-dsmz2}
\end{figure}

Neglecting the higher order corrections in wave vector in the Hamiltonian (\ref{lo}), describing 
Dirac semi-metal with two Dirac cones protected by the rotational symmetry of the crystal, one obtains
 \be
H=a(\tau^x\sigma^zk_x-\tau^yk_y)+b(\textbf{k})\tau^z.
\label{lowest}
\ee 
As the Hamiltonian in question commutes with spin operator $\sigma_z$, its eigenvalues can be labelled by the eigenvalues 
of $\sigma^z$, i.e., $s=\pm 1$. This fact enables us to conclude that the Hamiltonian for each spin projection implies
\be
H_s=a(s\tau^x k_x-\tau^yk_y)+b(\textbf{k})\tau^z,
\label{dsm-spin}
\ee 
being $2\times 2$ matrix.
Using the standard low energy form \cite{deng2019} for $b(\textbf{k})=m_0-m_1k_z^2-m_2(k_x^2+k_y^2)$ in the continuous limit one notices that $H_s$ for each spin eigenvalue $s=\uparrow, \downarrow$ contributes two Weyl nodes at $\textbf{k}_W^{C_2}=(0,0,C_{2}\sqrt{m_0/m_1})$, where $C_{2}=\pm 1$ denotes the $\zz$ charges of the Dirac points. Diagonalising Hamiltonian (\ref{dsm-spin}) one notes that the spectrum does nor depend on the spin quantum number $s$. To get the spectrum shown in Fig. (\ref{added-dsmz2}) we have put $k_x=k_y=0$ and expanded $b(\mathbf{k})$ near each of the nodes to the linear order in $k_z$.  One finds $E_{\pm}^{C_2}(0,0,k_z)=\mp 2\sqrt{m_0 m_1}(k_z-C_2\sqrt{m_0/m_1})$: the expression plotted in the figure for $m_0=m_1=1$. In equilibrium the nodes are at $E(0,0,k_z)=0$; the additional shifts mimic the $\zz$ chemical potential difference in non-equilibrium with $E(k_z)\rightarrow E_{C_2}(k_z)-\mu_{C_2}$.   

Taking the spin degeneracy into account, one finds in each of the nodes (differing by the value of $\zz$ charge)  fermions with spin up or down and with chirality ($+$) or ($-$).
The eigenvalues of the $\zz$ charge operator \cite{burkov2016, deng2019}
are  related to the chiralities $\chi_\pm=\pm 1$ of the spin up and down electrons by $C_{2}=(\chi_\uparrow-\chi_\downarrow)/2$. It follows that the node with $\zz$ charge $C_{2}=+1$ has spin and chirality $(\uparrow, +1)$,~ $(\downarrow, -1)$, while that with $C_{2}=-1$ is composed of fermions with (spin, chirality) quantum numbers $(\uparrow, -1)$,~$ (\downarrow, +1)$. 

Using the above phenomenological picture we propose the holographic description of the Dirac semimetal. To this end, and  in  analogy to the chirality in Weyl semimetals \cite{lan16, lan16a}
where the axial current $j_5$ is related to the axial charge $\rho_5$, to treat the corresponding 'axial part of the current between the nodes' we introduce two fields. The first denoted by $F_{\mu \nu}$ is standard Maxwell U(1)-gauge field and other, which we call $B_{\mu \nu}$ in the following plays the role analogous to $F^5_{\mu \nu}$ present in Weyl semimetals, where the charge and chirality are the only relevant quantum numbers. In principle, we could introduce $F^5_{\mu \nu}$ and and spin tensor field \cite{yu-chen2020} together with additional field playing the role of $F_{\mu \nu}$ in spin sector and use all four fields to calculate transport properties of the system. Such procedure, which would allow to calculate, e.g., inverse spin Hall effect is beyond the scope of the present paper.  Thus in the analysed Dirac semimetal, the field $B_{\mu \nu}$ is the analogue 
of the field $F^5_{\mu \nu}$ in the Weyl semimetals with axial current. 
However, in our approach the $\zz$ charge related current is induced by electromagnetic gauge field $F_{\mu \nu} $. 
The calculated below anomalous Hall effect is related to the spin degrees of freedom and is more properly called anomalous spin Hall effect. The word anomalous is related to the fact that it is intimately related to the vector \textbf{b} and not due to the magnetic field \textbf{B}, which also breaks time reversal symmetry. The magnetic field in Dirac semimetals with separated Dirac nodes is another source of the Hall effect and spin Hall effect, e.g., in the presence of spin dependent scatterings.  
 
In the presence of external  electric field \textbf{E} there is a shift of the nodes with a given $\zz$ charge in energy and with applied magnetic field  the spin current proportional to the \textbf{B} field is expected to appear in the system. The separation of the nodes in wave-vector space in the presence of the electric field contributes to the appearance of standard Drude current along the field and additional spin current perpendicular to the node separation vector \textbf{b}, unless \textbf{E} is along \textbf{b}.

In what follows we shall also focus on the role of the coupling constant between  fields $F_{\mu \nu}$ and $B_{\mu \nu}$. The coupling $\alpha$ between 
the aforementioned gauge fields, 
in the considered action (\ref{holact}), provides additional degree of freedom. Physically it could be related to the scattering processes.
 Contrary to the low-energy physics described by the free particle Hamiltonian, the holographic approach takes strong interactions into account.

The paper is organised as follows. In Section \ref{sec:model} we present the basic assumptions and equations of motion for the model in question. 
Section \ref{sec:3} is devoted to the Hall conductivity caused by the elaborated gauge fields.
In section \ref{sec:5} one derives the equations constituting the description of the longitudinal conductivity. Section \ref{sec:6} is dedicated to the numerical solutions of the underlying equations of motion, paying special attention to the role of $\alpha$-coupling constant and the influence of Chern-Simons terms on the physics of the studied phenomena. Section \ref{sec:7} concludes our studies. In the appendix we present some comments concerning the boundary currents in the underlying theory.

\section{Holographic model}
\label{sec:model} 
Topological semimetals being novel quantum states of matter can be classified in main two groups \cite{hsi08}-\cite{neu16}. The first one, in which the Dirac points fall out at time reversal invariant
momenta in the first Brillouin zone and the other one for which the Dirac points happen in pairs, separated in momentum space along a rotation axis. The latter one characterises by a non-trivial $\zz$ topological invariant, leading to the emergence of Fermi arc surface states, connecting projections of the node locations on the Brillouin surface. In \cite{burkov2016} it was revealed that this kind of Dirac  semimetals exhibited, in addition to the chiral the $\zz$ quantum anomaly. 
For the sake of completeness we add that each of the above groups of semimetals can break the Lorentz invariance and be further considered as being of type-I or type-II. Type-II semimetals are characterised by the over-tilted Dirac cones \cite{soluyanov2015}  and the existence of electron and hole pockets and thus finite density of states at the Dirac/Weyl point. In type I systems the cone  can also be tilted but the density of states vanishes at the node. Besides the above systems a special class of topologically nontrivial materials, so called nodal line semimetals \cite{burkov2011} exists, in which nodes appear along the closed line in the wave vector space.

The quantum model in question was described in terms of AdS/CFT correspondence \cite{rog19}, where the bulk action describing the system with chiral anomaly and $\zz$ topological charge was studied in the context of magnetic conductivity. The key point in our model are various combinations of gauge Chern-Simons terms mimicking the quantum properties of the system in question, i.e., relations among the indexes for spin, $\zz$ charge and chirality in topological semimetals.

The action is provided by
\ben \label{holact} \nonumber
S_{\zz} &=& \int dx^5 \sqrt{-g}\Bigg( R + \frac{12}{L^2} - \frac{1}{4}F_{\mu \nu} F^{\mu \nu} - \frac{1}{4}B_{\mu \nu} B^{\mu \nu} 
 - \frac{\alpha}{4} F_{\mu \nu} B^{\mu \nu} 
 + \frac{\alpha_1}{3}\ep^{\mu \nu \rho \delta \tau} A_\mu ~F_{\nu \rho}~F_{\delta \tau}\\ 
&+& \frac{\alpha_2}{3}\ep^{\mu \nu \rho \delta \tau} B_\mu ~B_{\nu \rho}~B_{\delta \tau}  + \frac{\alpha_3}{3}\ep^{\mu \nu \rho \delta \tau} A_\mu ~F_{\nu \rho}~B_{\delta \tau} 
+ \frac{\alpha_4}{3}\ep^{\mu \nu \rho \delta \tau} B_\mu ~F_{\nu \rho}~B_{\delta \tau} \Bigg). 
\een

The existence of the {\it kinetic} mixing term $\alpha$ has its origin in the cosmology where the two
fields are interpreted as {\it visible} and {\it dark} sectors. The cosmological data related to the abundance of visible and dark matter in the Universe constrain the value of $\alpha$ to be much less than unity ($\alpha \ll 1$). In the studies of thermal transport properties of graphene the two fields were introduced as a representation of two sorts of carriers existing at finite temperature at the charge neutrality point \cite{seo17} and shown to lead to quantitatively correct holographic description of the thermal conductivity of the graphene. The mixing between two fields studied in \cite{rog18} allowed for additional improvements. The condensed matter applications do not lead to such strong constraint on its value as cosmological arguments. However, even though the calculations show that $|\alpha|<2$ we generally consider here $0<\alpha \le 1$. We adopt this limitation in the present work.    
Moreover, it turns out that the conservation law of the vector
current requires that $\alpha_1=\alpha_3=0$ (see Appendix).

The transport properties of the model (\ref{holact}) have  been studied earlier using the hydrodynamic and holographic approaches \cite{rog18b,rog19}. We  paid special attention to the chiral anomaly and $\zz$ topological charge. The Chern-Simons parameters $\alpha_i$ have been shown to be directly related to the corresponding chiral anomaly parameters $C_i$ and to affect the magneto-transport characteristics of the material by introducing novel kinetic coefficients related, e.g., to the chiral magnetic and chiral vortical effects.  
The ability of the model (\ref{holact}) to provide the correct description of the magnetotransport of the Dirac system in the hydrodynamic regime \cite{rog18b,rog19} is the main motivation of its use to describe anomalous Hall conductivity in the holographic approach and the topological to trivial system phase transition with increasing the mass $M$.  Properties of the Dirac and Weyl semimetals have been
studied, for the non-interacting systems, by means of standard condensed matter techniques. On that level the  Berry phase and the topology of the 
Fermi surface are responsible for monopole like singularity of the Berry connection which mimics the effect of Chern-Simons terms and chirality \cite{son12}.


\subsection{Equations of motion}
The main objective of our paper is the action provided by
\be
S = S_{\zz} + \int dx^5 \sqrt{-g}\Bigg( - (D_\mu \Phi)^\dagger D^\mu \Phi - m^2 \Phi^\dagger \Phi \Bigg),
\ee
where the scalar field $\Phi$ appearing in (\ref{holact}) is charged under $B_\mu$ gauge field, i.e.,
\be
D_\mu \Phi = \Big( \p_\mu - i q_d B_\mu \Big) \Phi,
\ee
where $q_d$ is the charge connected with the auxiliary gauge field. The scalar mass is chosen in such way that it fulfils the Breitenlohner-Freedman limit, i.e., the bulk mass
satisfies the condition $m^2 = -3$.

In the holographic model the presence of gauge Chern-Simons terms (gauge curvature)  justifies the breakdown of $U(1)$-gauge symmetry.
The four different terms are forced by the charge and spin degrees of freedom in the considered $\zz$ Dirac semimetal. In the calculations below we keep all four coupling constants
$\alpha_i$ to have general non-zero values. However, in the Appendix we elaborate on the conservation 
of currents in the presence of anomaly and show that there  exist constraints on the 
allowed values of these parameters. 

We also consider the symmetry breaking by the non-zero mass term connected with a non-normalizable mode of the charged scalar field. 
It is worth to remark that the aforementioned influence of Chern-Simons term and symmetry breaking  by non-zero value of gauged scalar field was elaborated in \cite{ban14}.

In what follows, our convention is $\ep^{trxyz} =1$ and $\Phi=\phi(r)$. The equation of motion for the $U(1)$-gauge fields with scalar field charged under $B_\mu$ gauge
field can be written as
\ben
\na_\alpha F^{\alpha \beta} &+& 
\frac{\alpha}{2} \na_\alpha B^{\alpha \beta} +
\alpha_1~\ep^{\beta \mu \nu \rho \delta} F_{\mu \nu} F_{\rho \delta} + \frac{2}{3}\alpha_3~ \ep^{\beta \mu \nu  \rho \delta} F_{\mu \nu} B_{\rho \delta}\\ \nonumber
&+& \frac{\alpha_4}{3} ~\ep^{\beta \mu \nu \rho \delta} B_{\mu \nu} B_{\rho \delta} = 0.
\een
On the other hand, for the auxiliary $B_{\mu \nu}$ field strength one arrives at the relation
\ben
\na_\alpha B^{\alpha \beta} &+& 
\frac{\alpha}{2} \na_\alpha F^{\alpha \beta} +
\alpha_2~\ep^{\beta \mu \nu \rho \delta} B_{\mu \nu} B_{\rho \delta} + \frac{2}{3}\alpha_4 ~\ep^{\beta \mu \nu  \rho \delta} B_{\mu \nu} F_{\rho \delta} \\ \nonumber
&+& \frac{\alpha_3}{3} ~\ep^{\beta \mu \nu \rho \delta} F_{\mu \nu} F_{\rho \delta} 
- 2 q_d^2~\phi^2~B^\beta= 0.
\een
In the next step we can get rid of the terms with $\alpha$-coupling constant and finally get
\be
\na_\alpha F^{\alpha \beta} + 
\frac{\talpha_1}{\talpha}~\ep^{\beta \mu \nu \rho \delta} B_{\mu \nu} B_{\rho \delta} + \frac{\talpha_2}{\talpha}~\ep^{\beta \mu \nu \rho \delta} F_{\mu \nu} F_{\rho \delta}
+ \frac{\talpha_3}{\talpha}~ \ep^{\beta \mu \nu  \rho \delta} F_{\mu \nu} B_{\rho \delta} = 0,
\ee
where the coefficients $\talpha_i$ are given, respectively by
\be
\talpha_1 = \frac{\alpha_4}{3} - \frac{\alpha~\alpha_2}{6}, \qquad
\talpha_2 = \alpha_1 - \frac{\alpha~\alpha_3}{6}, \qquad
\talpha_3 = \frac{2}{3} \alpha_3 - \frac{\alpha~\alpha_4}{3}, \qquad
\talpha = 1 - \frac{\alpha^2}{4}.
\ee
As mentioned earlier the conservation law of the vector current requires $\alpha_1=\alpha_3=0$ (for the
detailed discussion of the constraints on coupling constants, from the point of view of gauge-current conservations, see Appendix).

The same procedure of obtaining equations of motion can be applied to the additional gauge field. It reveals the following relation:
\be
\na_\alpha B^{\alpha \beta} + 
\frac{\beta_1}{\talpha} ~\ep^{\beta \mu \nu \rho \delta} F_{\mu \nu} F_{\rho \delta} + \frac{\tbeta_2}{\talpha} ~\ep^{\beta \mu \nu \rho \delta} B_{\mu \nu} B_{\rho \delta} 
+ \frac{\tbeta_3}{\talpha} ~\ep^{\beta \mu \nu  \rho \delta} B_{\mu \nu} F_{\rho \delta} - 2 \frac{q_d^2~\phi^2~B^\beta}{\talpha}= 0, 
\ee
where $\tbeta_i$ read
\be
\tbeta_1 = \frac{\alpha_3}{3} - \frac{\alpha~\alpha_1}{2}, \qquad
\tbeta_2 = \alpha_2 - \frac{\alpha~\alpha_4}{6}, \qquad
\tbeta_3 = \frac{2}{3} \alpha_4 - \frac{\alpha~\alpha_3}{3}.
\ee 
The scalar field equation charged under $B_\mu$ field fulfils the following equation of motion:
\be
\na_\mu \na^\mu \phi - q_d^2~\phi~B_\mu B^\mu - m^2 \phi = 0.
\ee

In our consideration, as the background metric we take the line element of AdS-Schwarzschild five-dimensional black brane
\be
ds^2 = r^2 \Big( - f(r) dt^2 + dx^2 + dy^2 + dz^2 \Big) + \frac{dr^2}{r^2~f(r)},
\ee
where $f(r) = 1 - \frac{r_0^4}{r^4}$ and $r_0$ is the radius of the black brane event horizon, related to the Hawking temperature by $T=r_0/\pi$.  We shall work in the {\it probe limit}, neglecting the spacetime metric tensor
fluctuations.

\section{Anomalous spin Hall effect in the holographic Dirac semimetal}
\label{sec:3}

To commence with, in this section we consider the case when $B_\mu$ field will constitute the background field and one will elaborate the $A_\mu$ field as the fluctuations on the {\it spin}  $U(1)$-gauge field background. 
As in \cite{burkov2016}, we introduce the vector field (in the present notation $b_\mu$) which in the weak coupling description  couples to the Dirac fermions.
In what follows, without loss of generality we assume
\be
\vec b= b~ \vec e_z,
\ee
and correspondingly, in the holographic model, take the $z$-component ($B_z$) of the background field.
The background field equations of motion are provided by
\ben \label{az1}
B^{''}_z(r) &+& B^{'}_z(r)~\Bigg( \frac{3}{r} + \frac{f^{'}(r)}{f(r)} \Bigg) - 2 \frac{q_d^2~\phi^2(r)~B_z(r)}{ r^2~f(r)~\talpha} = 0, \label{eq:vis_bckgnd1} 
\\ 
\phi^{''}(r) &+& \phi^{'}(r)~\Bigg( \frac{5}{r} + \frac{f^{'}(r)}{f(r)} \Bigg)  - \phi(r) ~\Bigg(
\frac{q_d^2 B_z(r)^2}{r^4~f(r)} + \frac{m^2}{r^2~f(r)} \Bigg) = 0.
\label{eq:vis_bckgnd2}
\een
The boundary conditions are given by the Dirac cones separation parameter $b$ and the fermion mass $M$ parameter, respectively
\be
\lim_{r \rightarrow \infty} B_z(r) = b, \qquad \lim_{r \rightarrow \infty} r~\phi(r) = M. 
\label{eq:bc_vis}
\ee

Note, that in this respect both Weyl and Dirac systems are not distinguishable,  both the nodes are separated by $b$.

\subsection{Hall conductivity}
Our main task will be to find the Hall conductivity, using  Kubo formula given by 
$\sigma_{xy} = \lim_{r \rightarrow \infty} \frac{1}{i \omega} \langle J_x, J_y \rangle_{ret}(\omega, k=0)$. 
In the holographic approach the Green function can be found by studying fluctuations of the gauge fields, dual to the currents, around the considered background line element, with respect to the in-falling boundary conditions at the event horizon of the studied black brane spacetime. Namely, one finds
the mode equations with the solutions equal to 1 at the boundary. For timelike momenta the solution should have asymptotic expression
envisaging the incoming wave at the event horizon, while for spacelike momenta the solution ought to be regular at the horizon. Only contributions from the 
boundary are taken into account, surface terms corresponding to the event horizon are dropped out. This part of the metric influences the Green function in question only by boundary conditions imposed on the bulk fields \cite{son02}-\cite{lan16acta}.

As in \cite{lan16} the retarded correlation function can be calculated having in mind the fluctuations of the adequate fields above given background. Let us
consider Maxwell $U(1)$-gauge field fluctuations provided by the following:
\be
\delta A_x = a_x(r) e^{-i \omega t}, \qquad \delta A_y = a_y(r) e^{-i \omega t}.
\ee
In the next step one defines the quantity binding $b_x(r)$ and $b_y(r)$ in the form as
\be
a_{\pm} (r) = a_x(r)  \pm i~a_y(r).
\ee

It leads to the following form of the underlying equations:
\be
a_{\pm}^{''}(r) + a_{\pm}~\Bigg( \frac{3}{r} + \frac{f^{'}(r)}{f(r)} \Bigg)   + \frac{\omega^2}{r^4~f(r)^2} ~a_{\pm}(r) \pm
4 \frac{\talpha_3 ~\omega}{\talpha} \frac{ B^{'}_z(r)}{r^3~f(r)}~a_\pm(r) = 0.
\ee
After a convenient parameterisation
\be
a_{\pm} = f(r) e^{- \frac{i \omega}{4 r_0}}~\Big( a_\pm^{(0)} + \omega a_\pm^{(1)} + \dots \Big),
\ee
we obtain the relations for zero and first order expansions in $\omega$
\ben \label{b0}
a_\pm^{(0) ''} &+& \Bigg( \frac{3}{r} + \frac{f^{'}(r)}{f(r)} \Bigg) ~a_\pm^{(0) '} = 0, \\ \label{b1}
 a_\pm^{(1)''} &+& \Bigg( \frac{3}{r} + \frac{f^{'}(r)}{f(r)} \Bigg) ~a_\pm^{(1) '}  = \frac{i}{4 r_0} ~\Bigg(
 \frac{3~f(r)^{'}}{r~f(r)} + \frac{f^{''}(r)}{f(r)} \Bigg)~a_\pm^{(0)} \\ \nonumber
 &+& \frac{i}{2 r_0} ~\frac{f(r)^{'}}{f(r)}~a_\pm^{(0)'}
 \mp 4 \frac{\talpha_3 ~\omega}{\talpha} \frac{ B^{'}_z(r)}{r^3~f(r)}~a_\pm^{(0)}. 
 \een
The regularity condition near the black brane event horizon leads to the conclusion that the solution of the equation (\ref{b0})
is given by $a_\pm^{(0)} = a_0$, where $a_0$ is a constant. On the other hand, the integral constant method applied to the relation (\ref{b1})
implies
\be
a_\pm^{(1)}  = - \int_{r_o}^{\infty} dy~\frac{a_0}{y^3~f(y)}~\Big[
\frac{i~y^3~f^{'}(y)}{4 r_0} - i r_0 \mp \frac{4 \talpha_3}{\talpha}~\Big(B_z(y) - B_z (r_0) \Big) \Big].
\ee
The Green function implies
\be 
G_\pm (r)= \omega~\Big[ i r_0 \pm \frac{4 \talpha_3}{\talpha}~\Big(b - B_z (r_0) \Big) \Big].
\label{ga}
\ee

Having in mind relation (\ref{ga}), one gets that
\be
\sigma_{xy}= \frac{G_+- G_-}{2 \omega} = \frac{4 \talpha_3}{\talpha}~\Big(b - B_z (r_0) \Big).
\label{halla}
\ee
The first term in equation (\ref{halla}) originates from the Chern-Simons gauge term contribution to the currents (it constitutes the Bardeen-Zumino like polynomial contribution). 
In order to obtain the correct charge conserving definition of the current and accurate value of the appropriate kinetic coefficient, one ought to subtract it \cite{lan16acta}. 
On this account, at leading order in $\omega$, we achieve the following relation:
\be
\sigma_{Hall}= \frac{4 \talpha_3}{\talpha}~B_z(r_0).
\label{hallanom}
\ee
We remark here that the general value of $\talpha_3$ found earlier, reduces to $\talpha_3 =  - \frac{\alpha~\alpha_4}{3}$ if the constraints found in Appendix are taken into account.
On the other hand, the conductivities in $x$ and $y$-directions imply respectively
\be
\sigma_{xx}= \sigma_{yy} =  r_0.
\ee


\section{Longitudinal conductivity}
\label{sec:5}
This section will be devoted to the longitudinal electric conductivity at zero density. We shall consider 
the fluctuation in the background, which does not source other modes at zero density.
The fluctuation of the {\it charge } gauge field is of the form
\be
\delta A_z = \nu_z(r) e^{- i \omega t},
\ee
\be
\nu_z^{''}(r) + \nu_z^{'}(r) ~\Bigg( \frac{3}{r} + \frac{f^{'}(r)}{f(r)} \Bigg) + \frac{\omega^2~\nu_z(r)}{r^4~f^2(r)} = 0.
\ee
The same consideration as in the previous sections lead to the conclusion that in the $\omega^0$ order $\nu_z^{(0)} = k_0$, where $k_0$ is constant.
On the other hand, in $\omega^1$-order we have
\be
\nu_z^{(1)''} + \Bigg( \frac{3}{r} + \frac{f^{'}(r)}{f(r)} \Bigg) ~\nu_z^{(1) '}  = \frac{i}{4 r_0} ~\Bigg(
 \frac{3~f^{'}(r)}{r~f(r)} + \frac{f^{''}(r)}{f(r)} \Bigg)~\nu_z^{(0)} 
 + \frac{i}{2 r_0} ~\frac{f^{'}(r)}{f(r)}~\nu_z^{(0)'}.
 \ee
Thus the solution is provided by
\be
\nu_z^{(1)}  = - \int_{r_o}^{\infty} dx~\frac{k_0}{x^3~f(x)}~\Bigg(
\frac{i~x^3~f^{'}(x)}{4 r_0} - i r_0 \Bigg).
\ee
The form of the above relation implies that the component of the conductivity $\sigma_{zz} =r_0$. Its value does not depend on the considered gauge field, but
on the background geometry. Namely, the radius of the black brane event horizon.


\section{Numerical results}
\label{sec:6}
Let us now turn our attention towards the numerical solutions of the equations that had been derived in the former sections. 
The background {\it spin} $U(1)$-gauge field condensation leads to  two coupled ordinary differential equations.
We solve them by virtue of a shooting method, with the adequate boundary conditions (\ref{eq:bc_vis}) to be fulfilled on the $r \rightarrow \infty$ boundary.
The initial values of the fields $\phi(r_0)$ and $B_z(r_0)$ are our shooting parameters, moreover their derivatives are specified by the equations of motion.
Therefore our boundary value problem is translated into the initial value problem with the additional constraint equations on $r \rightarrow \infty$ boundary. These  
constitute a simple pair of algebraic equations, with the initial field values as variables.
The resulting equations are sequentially integrated with a standard Runge-Kutta-type method, with initial values correction in each step, until the satisfying precision is achieved. We have used Newton-Raphson method for minimising the conditions on the AdS spacetime boundary. 

\begin{figure}[h!]
 \centering
    \subfloat[Scalar field]{{\includegraphics[width=0.45\textwidth]{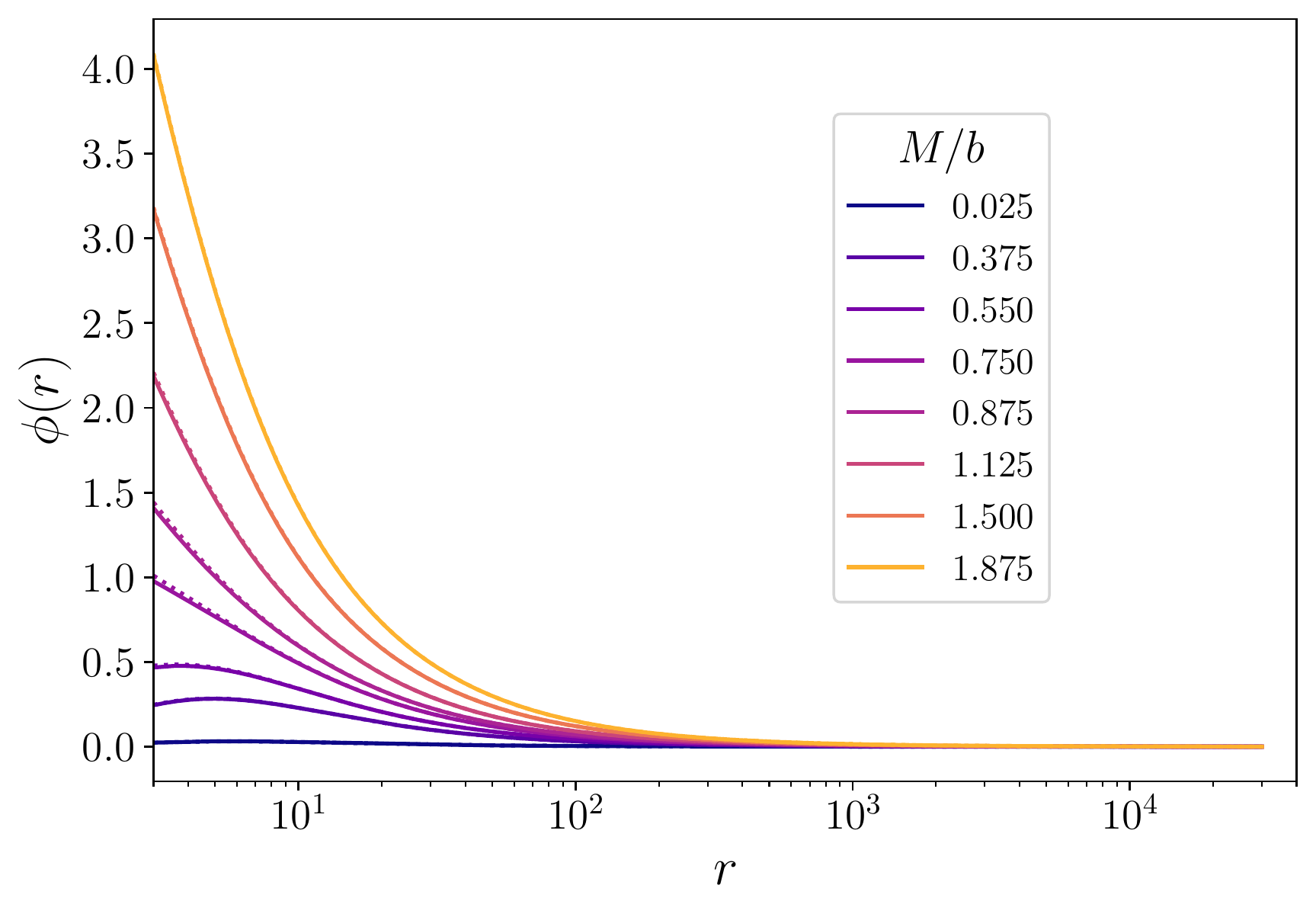} }}%
    \qquad
    \subfloat[The gauge field z-component]{{\includegraphics[width=0.45\textwidth]{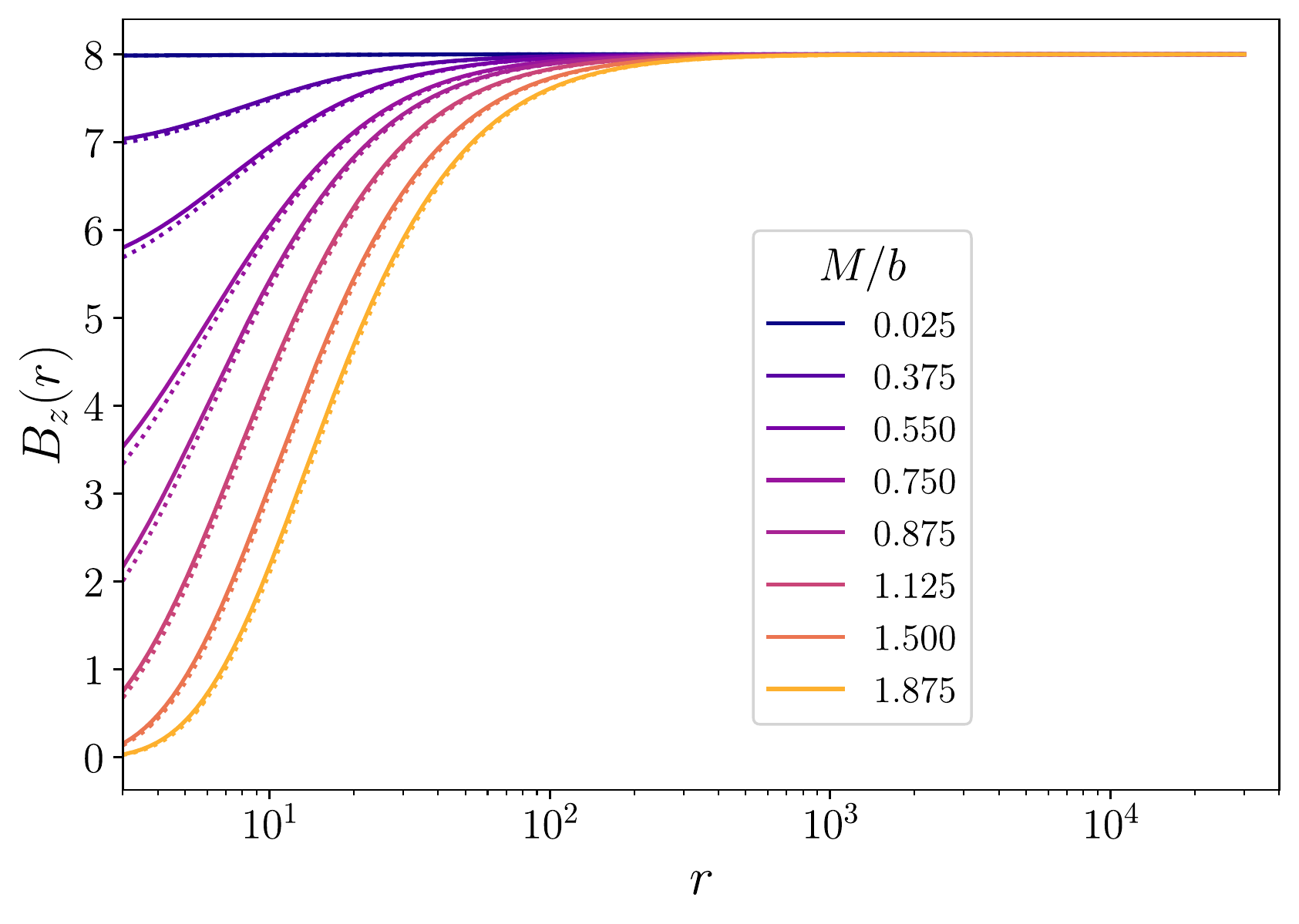} }}%
    \caption{Background fields as a function of renormalisation group flow coordinate. Solid lines represent solutions with $\alpha$-coupling constant equals to zero, 
    while dotted ones illustrate $\alpha = 0.4$ case.}%
    \label{fig:alphabeta_maps1}%
\end{figure}

In Fig. \ref{fig:alphabeta_maps1} we depicted the solutions of the equations (\ref{az1})-(\ref{eq:vis_bckgnd2}) for the scalar field $\phi$ and the background gauge $B_z$ as functions 
of renormalisation group flow coordinate, i.e., the radial coordinate $r$ of our AdS spacetime. For the coupling constant $\alpha =0$, they are given by solid lines, while for $\alpha = 0.4$ by dotted lines.
We can observe that scalar field solutions for $\alpha =0$ are slightly  below the ones with $\alpha \ne 0$. On the contrary, $B_z$ solutions, responsible for the conductivity, with $\alpha =0$ are visually hardly distinguishable from 
 the solutions with $\alpha \ne 0$. 
 
As the holographic direction can be understood as changing the energy scale, the obtained profiles for $B_z(r)$ and $\phi(r)$ illustrate the changes of the fields due to the alternation of energy scale from low to high values. The detailed behaviour depends on the ratio $M/b$. For small ratios $M/b$ the scalar field first starts growing with growing $r$  
then there is a value of $r$-coordinate for which the situation changes and its value decreases  to zero. For the values of $M/b$ bigger then critical one (see below) we observe the continuous decrease  of the scalar field towards zero value with increasing $r$ in such a way that the boundary condition (\ref{eq:bc_vis}) is satisfied. 
The bigger is the ratio $M/b$, the lager maximal value of
$\phi(r)$  for $r$ close to the horizon one achieves.

On the other hand, the gauge field  envisages the monotonic growth tendency, i.e., $B_z(r)$ is growing towards its limiting value $b$ as $r\rightarrow \infty$. 
The bigger $M/b$ is taken into account, the smaller value of $B_z(r)$ deep in the interior of AdS spacetime one obtains.
Next, for all the studied cases of $M/b$, they attain the same UV value. The UV limit is obtained earlier for smaller $M/b$ ratios.  As in \cite{lan16}, the limiting $B_z(r)$ value
is connected with the Hall conductivity and it authorises the holographic analogue of $b_{eff}$.


\subsection{Influence of $\alpha$ coupling} \label{sec:num_alpha}

\begin{figure}[h!]
\centering
\includegraphics[width=0.7\textwidth]{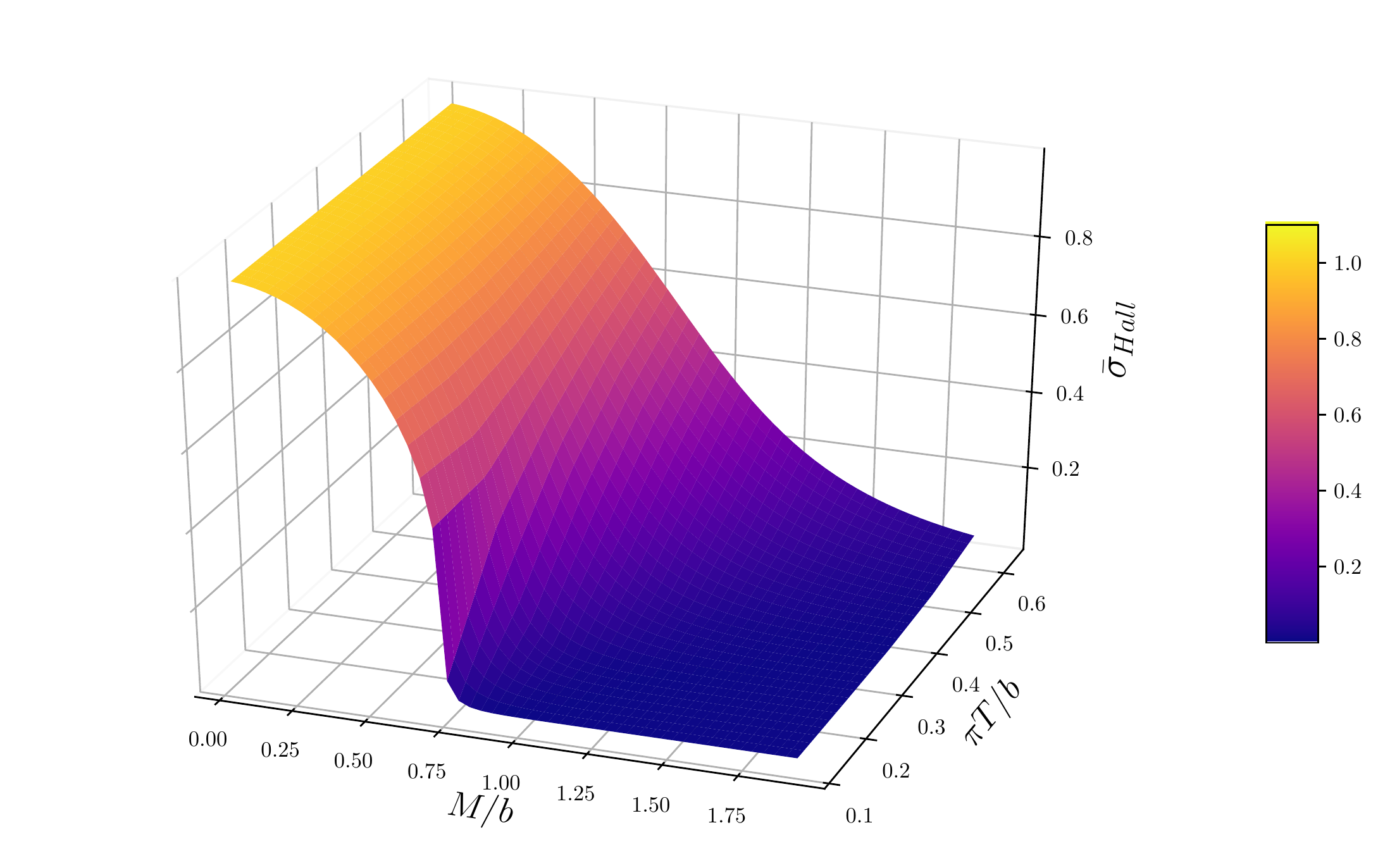}
\caption{Normalised anomalous Hall conductivity $\bar{\sigma}_{Hall}= \tilde{\alpha} \sigma_{Hall} / 4 \talpha_3 b$ from the relation \eqref{hallanom} as a surface plot with mass $M$ and temperature $T$ as variables. We set $b = 8$ and $\alpha = 0$.}
\label{fig:anom_hall3d}
\end{figure}

In Fig. \ref{fig:anom_hall3d} we plot the normalised
anomalous spin Hall conductivity $\bar{\sigma}_{Hall} = \tilde{\alpha} \sigma_{Hall} / 4 \talpha_3 b $, calculated from equation \eqref{hallanom}, as a function of fermionic mass $M$  and temperature $T$. The normalisation makes $\bar{\sigma}_{Hall}$ a dimensionless quantity. This normalisation is convenient as it does not require fixing the values of the couplings constants of various Chern-Simons terms. 

The non-zero Hall conductivity corresponds to the topologically non-trivial phase of the system and thus serves as an order parameter of the considered phase transition. By increasing the mass $M$ one drives the system towards a quantum phase transition into a topologically trivial phase. 
For low temperatures the characteristics are steep and conductivity drops to zero around $M/b \approx 0.72$. 
If one increases the temperature the phase boundary is not so sharp any more and the conductivity profile stretches. 
Moreover, at elevated temperatures the non-trivial topological state persists to larger values of mass.
In the particular case of the Fig. \ref{fig:anom_hall3d} we use $b = 8$ and $\alpha = 0$. We do not plot surfaces for alternative $\alpha$-coupling values, because they would be hardly distinguishable.

\begin{figure}[h]
\centering
\includegraphics[width=0.7\textwidth]{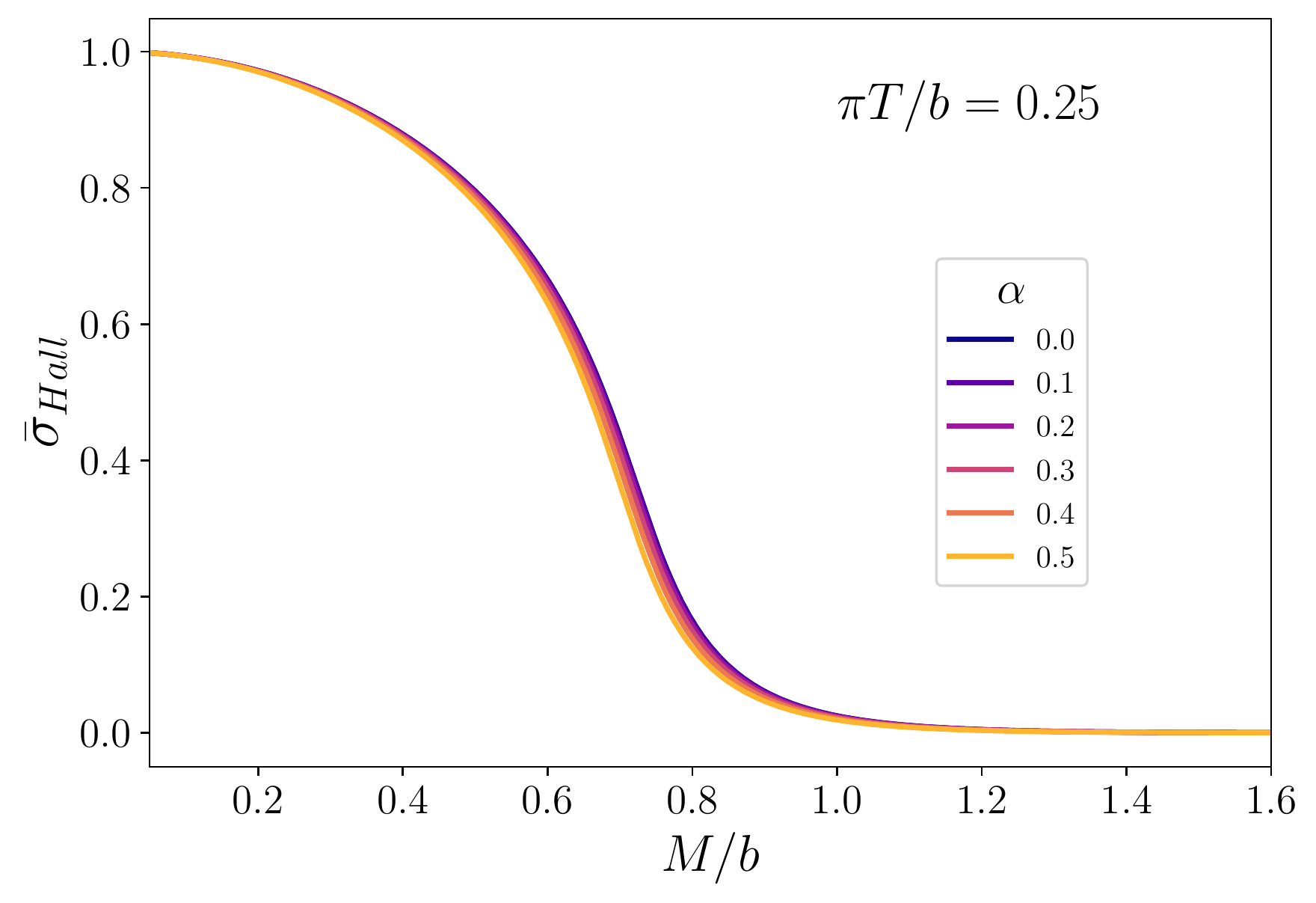}
\caption{Normalized anomalous Hall conductivity profile versus fermionic mass for different values of $\alpha$-coupling.
The figure presents an exemplary slice for constant temperature from Fig. \ref{fig:anom_hall3d}. One can notice that kinetic mixing term decreases the conductivity in topologically non-trivial phase, especially in the steep central region of the characteristics.
}
\label{fig:anom_hall_slice}
\end{figure}

Fig. \ref{fig:anom_hall_slice} presents such dependence. The normalised Hall conductivity is plotted as a function of mass with constant temperature, for different values of $\alpha$ parameter.
It can be seen that the presence of the kinetic mixing term causes the decrease of the critical point $(M/b)_{crit}$. Also the conductivity slightly decreases, especially in the steep region of the curve, in the vicinity of the quantum phase transition point.
This effect is subtle but observable. It can constitute the key point in identifying the existence of $\alpha$-coupling constant between two sectors of $U(1)$-gauge fields.

At this point it will be instructive to examine
more carefully the transport coefficient, at lower temperature, as well as, its dependence on the relative ratio of the mass $M$ and temperature $T$. As we have alluded in the introductory discussion the mass $M$ and the parameter $b$ define a gap in the Dirac semimetal spectrum. 
 It constitutes the relation between the gap in the spectrum and temperature, which affects the observation of the true phase transition. The smooth behaviour of the Hall conductivity at high temperatures has  already been observed in Fig. \ref{fig:anom_hall3d}.

\begin{figure}[h]
\centering
\includegraphics[width=0.75\textwidth]{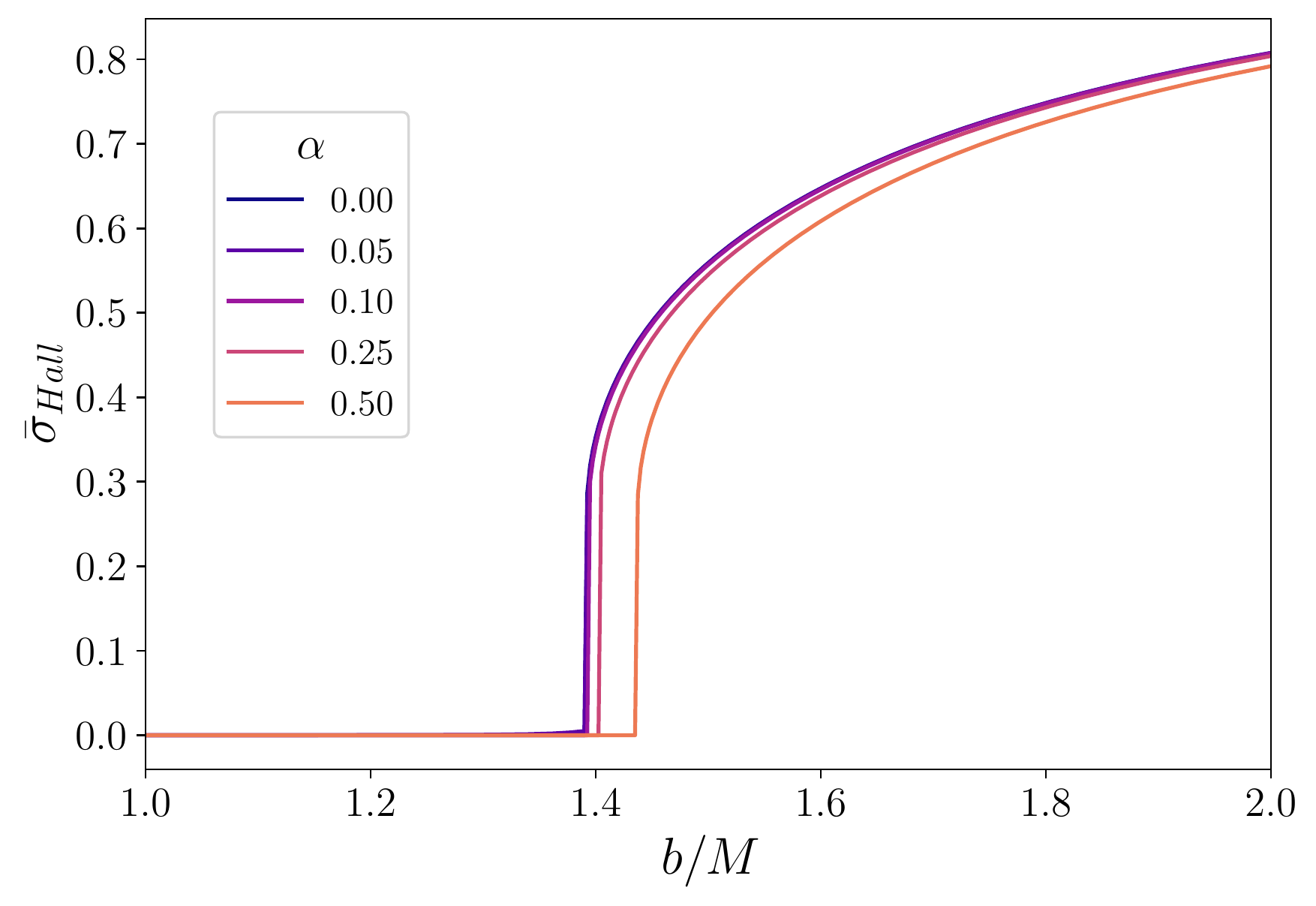}
\caption{Anomalous Hall conductivity as a function $b/M$ scale calculated for temperature $\pi T/M = 0.1$. The phase transition is quite sharp, with a delicate tail for $b/M < (b/M)_{crit}$. The presence of $\alpha$-coupling shifts the critical value of the holographic scale parameter.}
\label{fig:sigma_bm}
\end{figure}

To commence with, let us investigate the critical behaviour of the system near the quantum phase transition point, driven by the change of $b/M$ scale.
As one goes down with the temperature, the smeared tail of Hall conductivity (as it is seen in Fig. \ref{fig:anom_hall_slice}) vanishes and the phase transition becomes sharp. Naturally in our theoretical set up we cannot achieve the exact zero temperature, as we work in the probe limit and use the gravitational background of a black brane 
with defined Hawking temperature instead of a 
gravitational soliton metric.
Nevertheless we can lower the temperature to the point that allows us the  approximate analysis of the critical behaviour of the order parameter. To obtain quantitative information we fit the anomalous Hall conductivity close to the critical point by a power law in the form
\begin{equation}
\sigma \sim (b/M - (b/M)_{crit})^{\beta}.
\end{equation}
Similar analysis for the closely related model has been already done  in \cite{ammon2018} in the context of holographic disordered Weyl semimetal. However, the presence of $\alpha$ coupling is interesting from the perspective of the phase transition analysis in the Dirac semimetal. The results  are presented in Fig. \ref{fig:sigma_bm}.
It can be clearly seen that the quantum phase transition point $(b/M)_{crit}$ is influenced by $\alpha$. The value of $(b/M)_{crit}$ clearly grows as the coupling increases. 

The value of the critical exponent $\beta \approx 0.213$, calculated at temperature $\pi T/M= 0.1$, is in quite good agreement with the previously obtained zero temperature result $\beta\approx 0.211$ \cite{ammon2018,lan16a}. It has to be mentioned that the field theory model predicts standard mean field value $\beta=0.5$. We shall also like to add that in the 
present approach all other components of conductivity tensor are given by $r_0\propto T$.

\begin{figure}[h]
\centering
\includegraphics[width=0.75\textwidth]{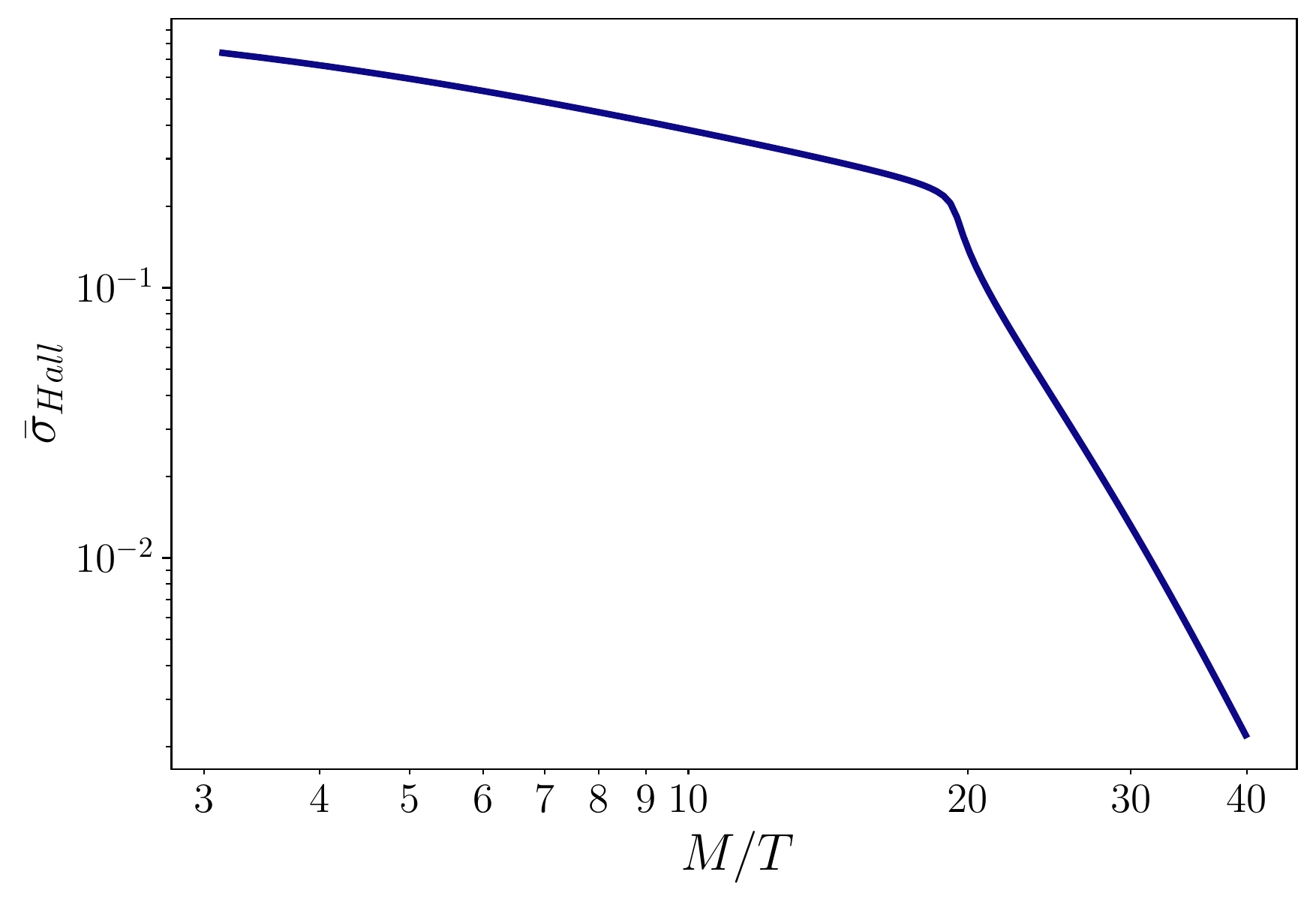}
\caption{A Log-Log plot of anomalous Hall conductivity versus $M/T$ scale calculated for the  value  $b/M\approx 1.399$, being very near the critical value, for a given $\alpha$. One can see two different behaviours in the quantum critical region, a power law function for small $M/T$, i.e., high temperature and exponential dependence for higher masses or lower temperatures. This behaviour is unaffected by the $\alpha$-coupling parameter, which only enters the position of the critical $(b/M)_{crit}$.}
\label{fig:sigma_mt}
\end{figure}

We also investigated the anomalous Hall  conductivity as a function of $M/T$ scale. Interestingly it features two different behaviours. For small values of $M/T$ the function follows the power law, while for the bigger values of the mass (or smaller temperature) it behaves exponentially, see Fig. \ref{fig:sigma_mt}. It has to be noted that similar behaviour has been observed by Ammon and co-workers \cite{ammon2018}. 
This scaling dependence is observed at, or in the vicinity of the critical point. The coupling $\alpha$ mainly changes the position of the critical point, as visible in Fig. \ref{fig:sigma_bm}. This explains our numerical observation in Fig. \ref{fig:sigma_mt} that the  scaling of the Hall conductivity is essentially independent on the coupling 
constant $\alpha$.

In the next step, we focus on the properties of Hall conductivity calculated using
the equations \eqref{eq:vis_bckgnd1} and \eqref{eq:vis_bckgnd2}.
To begin with, 
let us first discuss the behaviour of $\sigma_{xy}$, given by equations (\ref{halla}). We
take $b = 8$ and $q_d = 1$, and ignore the constant factor originating from Chern-Simons couplings, considering the normalised conductivity on its own.
In this way we can extract the influence of $\alpha$ coupling on the holographic Hall conductivity as a function of the fermionic mass to the gap ratio.

\begin{figure}[h]
\centering
\includegraphics[width=0.75\textwidth]{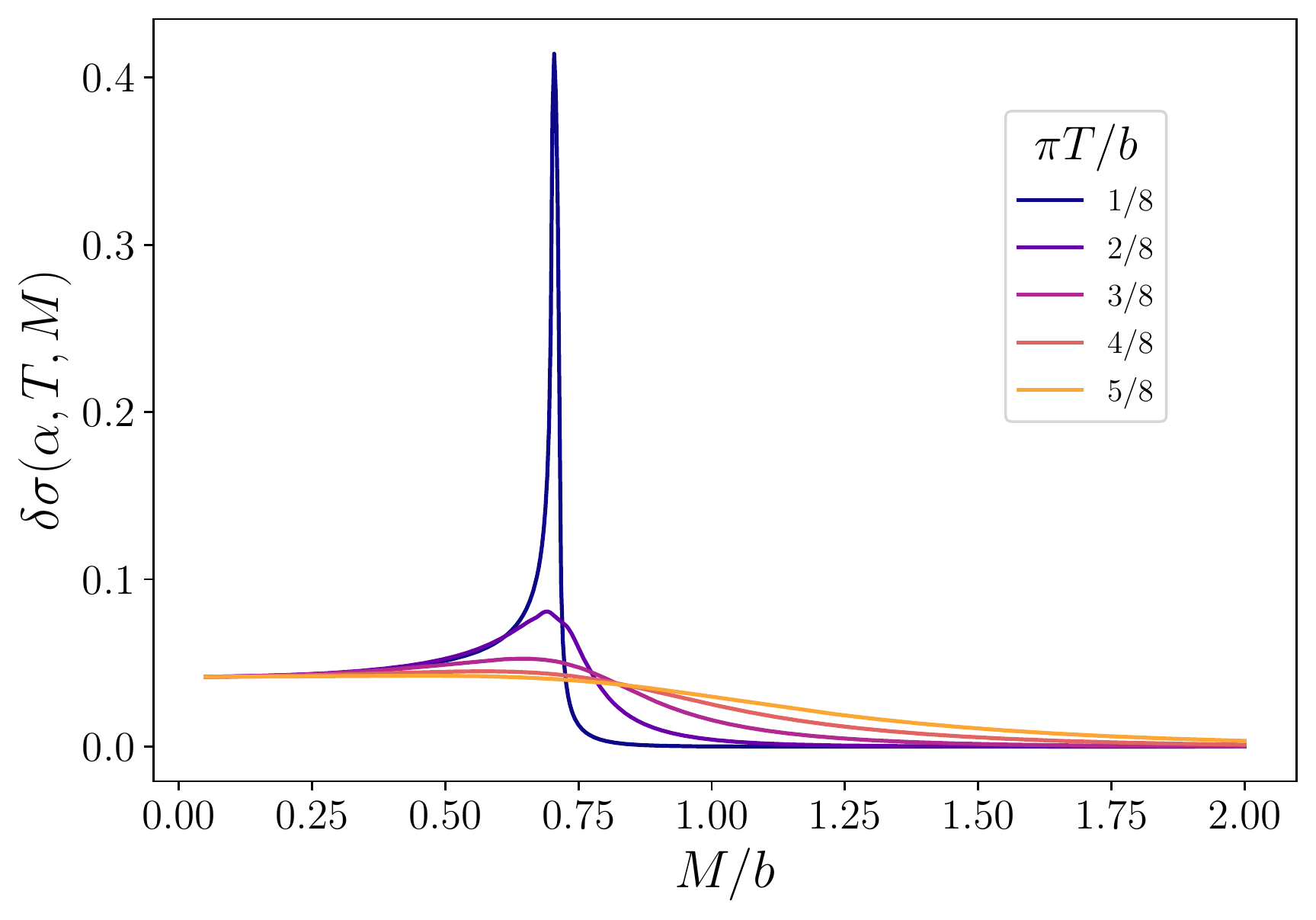}
\caption{The conductivity shift ratio defined in equation (\ref{deltasigma}) as a function of normalised fermionic mass, calculated for a number of temperature values. The coupling parameter is fixed to $\alpha = 0.4$. The function is strongly peaked in the low temperature around the point $M/b \approx 0.715$.}
\label{fig:dsigma_tshift}
\end{figure}

Consequently, let us
consider the following  dimensionless ratio:
\begin{equation}
\delta \sigma(\alpha, T, M) = \frac{\sigma_{xy}^{i}(\alpha, T, M) - \sigma_{xy}^{i}(0, T, M)}{\sigma_{xy}^{i}(0, T, M)},
\label{deltasigma}
\end{equation}
dependent  on the  coupling parameter $\alpha$, temperature $T$ and fermion mass $M$.
For simplicity we omit the indices, thus $\delta\sigma(\alpha, T, M) = \delta\sigma_{xy}(\alpha, T, M)$ in the text and in figures.
The following two figures present Hall conductivity ratio $\delta \sigma(\alpha, T, M)$ for two different scenarios. 
In the first one, presented in Fig. \ref{fig:dsigma_tshift}, we fix the $\alpha$-coupling constant to $\alpha = 0.4$ and increase Hawking temperature.
It can be observed that for low temperatures the coupling constant causes a strongly peaked shift in the Hall conductivity around a specific mass to gap ratio. With increasing temperature the peak blurs away due to thermal fluctuations. Apparently, the position of this peak coincides with the point of quantum phase transition, observed here, at finite but low enough  temperature, see Fig. \ref{fig:anom_hall3d}, \ref{fig:anom_hall_slice} and especially \ref{fig:sigma_bm}.
The imprints of the interactions  between the considered two $U(1)$-gauge sectors, mediated by $\alpha$, are the strongest in the vicinity of the critical point.  

\begin{figure}[h]
 \centering
    \subfloat[$\pi T/b = 1/8$]{{\includegraphics[width=0.45\textwidth]{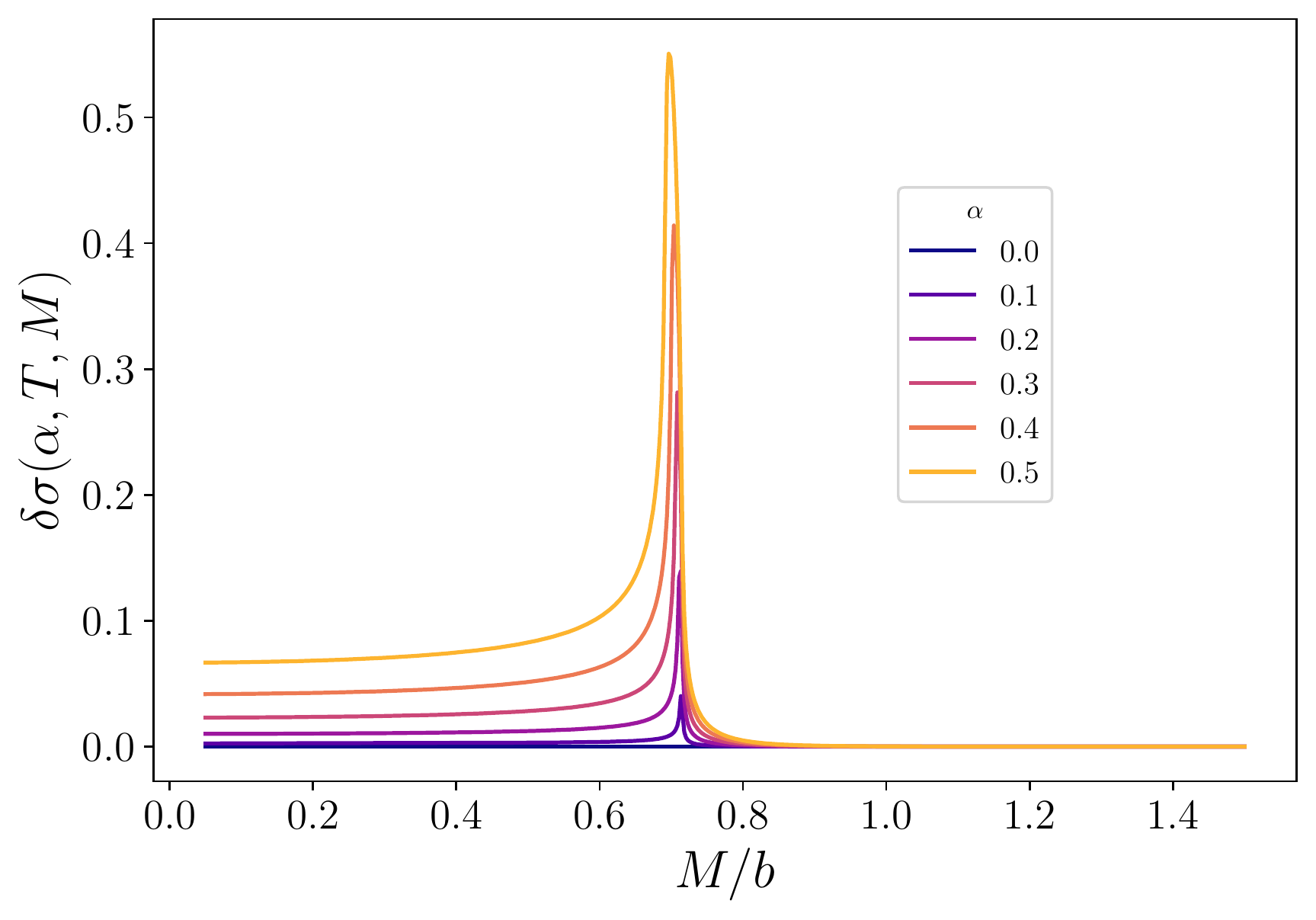} }}%
    \qquad
    \subfloat[$\pi T/b = 3/8$]{{\includegraphics[width=0.45\textwidth]{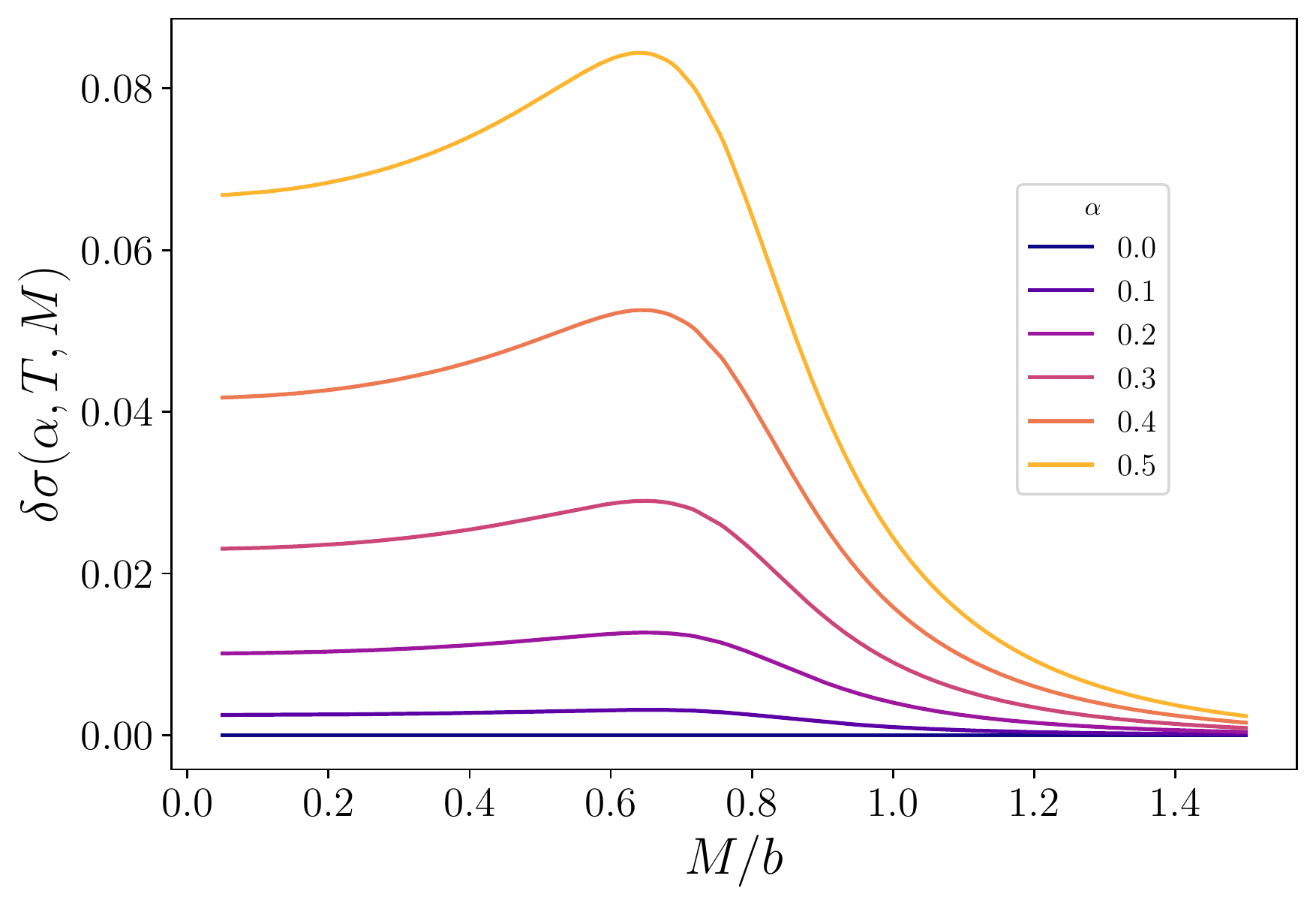} }}%
\caption{The ratio $\delta \sigma$ as a function of mass parameter $M/b$  calculated for a number of couplings $\alpha$. Both panels present the same function although for different temperatures. In the panel (a) the conductivity ratio at $\pi T/b=1/8$ and for $\alpha=0$ is strongly peaked around $M/b \approx 0.715$. The holographic approach supports an intuitive conclusion, that the rise of temperature smears the peak, as visible from panel (b) which has been obtained for $\pi T/b=3/8$}%
\label{fig:dsigma_ashift}%
\end{figure}

On the other hand, one may fix the temperature and vary the coupling strength parameter, which shows us the similar picture seen from a different perspective.
Fig. \ref{fig:dsigma_ashift} presents the conductivity shift ratio as a function of $M/b$. Once again we can see that it is peaked near specific values of $M/b$  with the peak position and its magnitude visibly dependent on $\alpha$. 
Namely, at this point one has that the larger value of $\alpha$-coupling constant we take into account, the bigger $\delta \sigma$  one achieves.
It is worth to recall that the non-interacting system is expected to have the critical value of $M/b=1$. 
This large renormalisation  of the critical value is attributed to strong coupling effects taken into account by holographic approach. 
It is also worth to note, that similarly large renormalisation of the critical value of the mass has been achieved in the lattice model, where it can be attributed to non-trivial energy spectrum. The critical exponents, however remain of the mean field variety, contrary to their holographic values.
 For the higher values of the temperature (right panel of the Fig. \ref{fig:dsigma_ashift}) the transition changes into crossover which is not so sharp and $\delta \sigma$ takes on small but non-zero values for $M/b$ well above 1. 

\begin{figure}[h]
\centering
\includegraphics[width=0.75\textwidth]{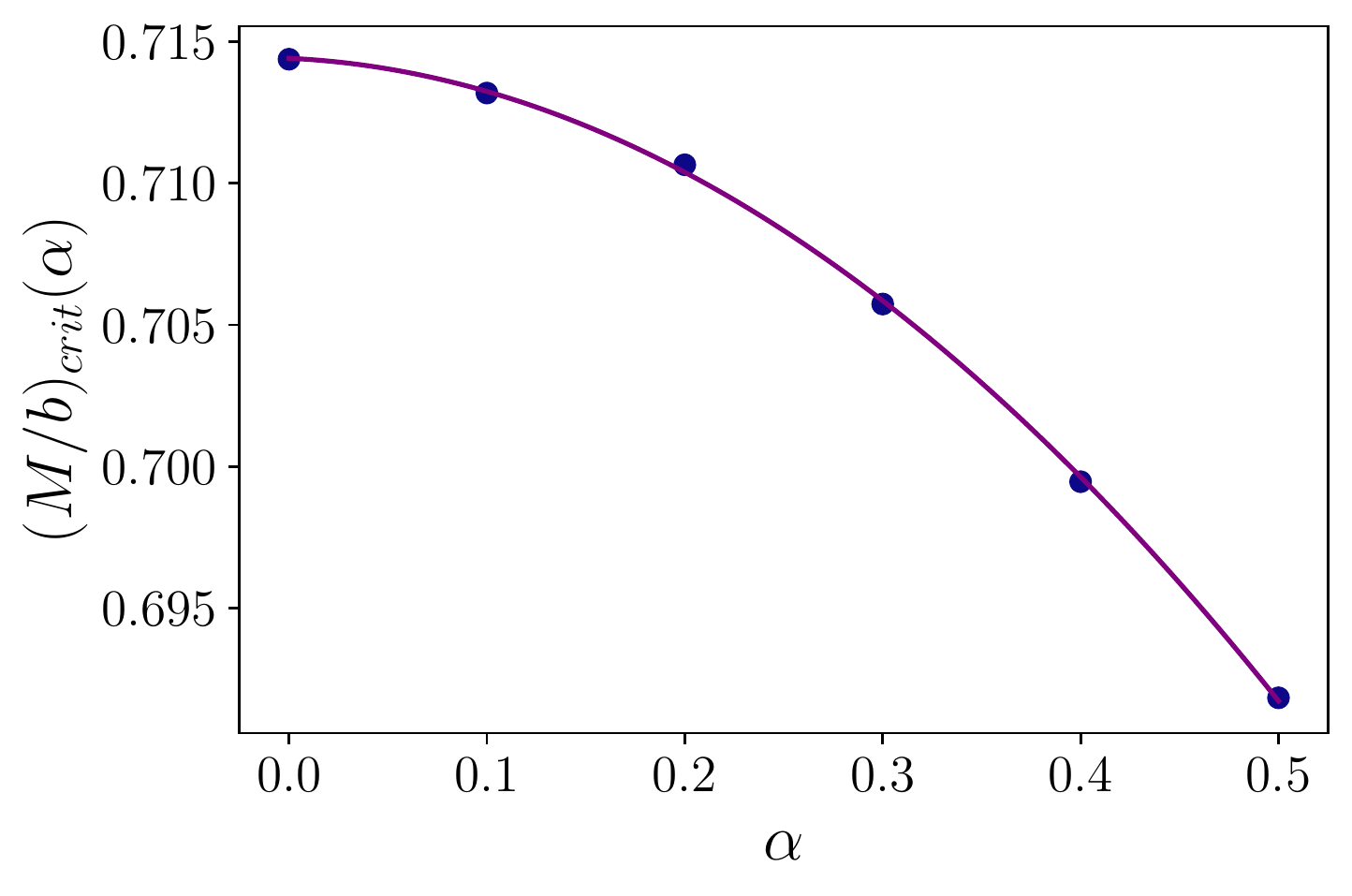}
\caption{The dependence of the critical value of $M/b$ on $\alpha$. The points are the numerical values deduced from the low temperature plots similar to that in Fig. \ref{fig:anom_hall_slice}, while the continuous curve is a parabolic fit to the points. }
\label{fig:mcrit_alpha_t18}
\end{figure}

\begin{figure}[h]
\centering
\includegraphics[width=0.7\textwidth]{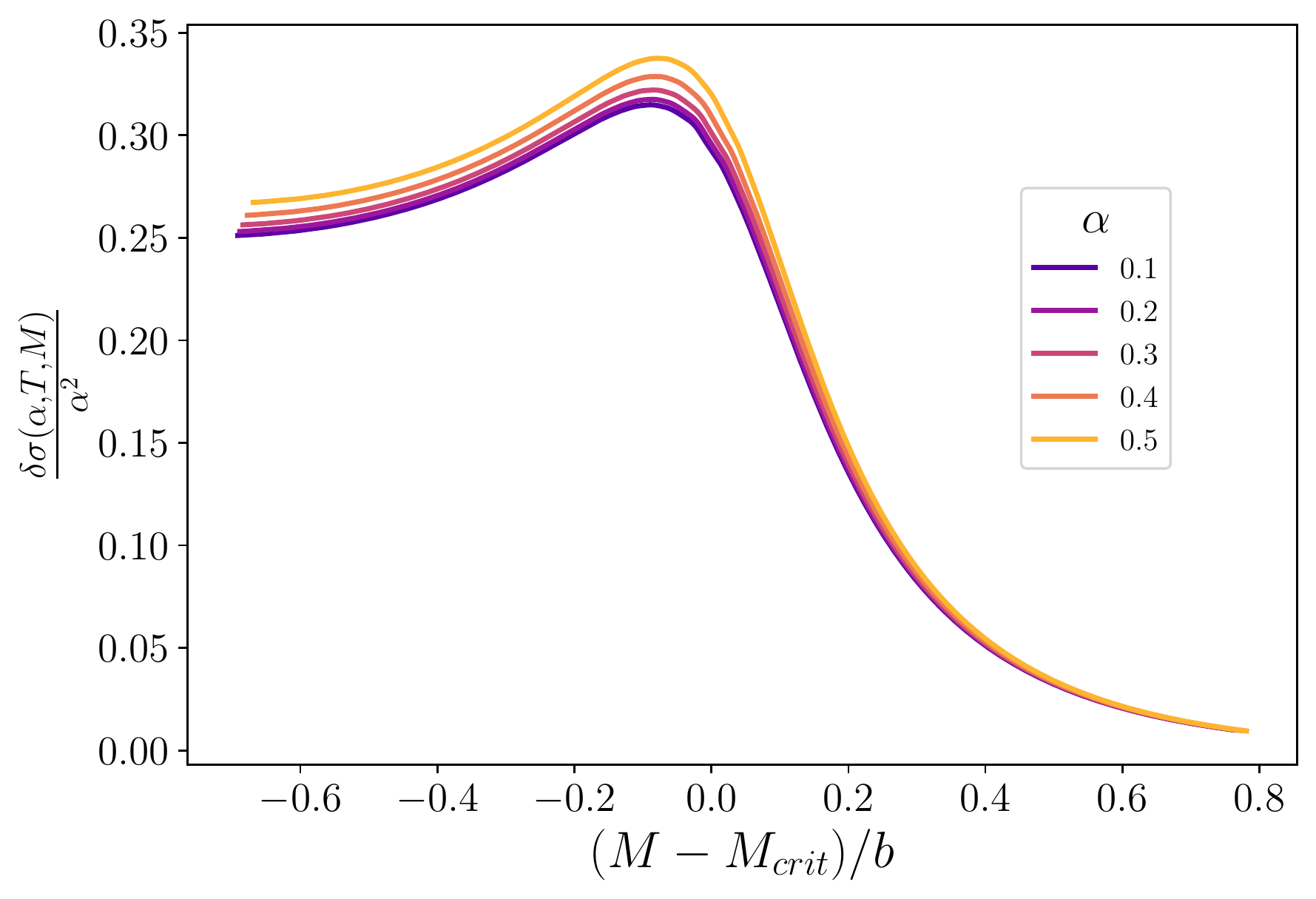}
\caption{The data from Fig. \ref{fig:dsigma_ashift}b rescaled according to the equation \eqref{eq:scaling}, with $n=2$. The rescaled conductivity ratios tend to one curve. However, the simple integer power $n=2$ does not lead to the collapse of data onto the single curve.
A slightly different power $2.03 < n < 2.04$ leads to much better convergence of the curves, nevertheless it is still not ideal.}
\label{fig:scaled_curves}
\end{figure}

It is well visible in Fig. \ref{fig:dsigma_ashift} that both the $\delta \sigma$ and the critical value of $M/b$ depend on the coupling $\alpha$. To learn more about this dependence we first numerically establish the dependence $(M/b)_{cr}(\alpha)$. For this we have defined these critical values from the low temperature plots of $\sigma_{Hall}$ as function of $M/b$,  similar to those shown in Fig. \ref{fig:anom_hall_slice}. The numerical points  and the parabolic fit to them are shown in Fig. \ref{fig:mcrit_alpha_t18}. The dependence
is approximately quadratic with $(M/b)_{crit}(\alpha)=(M/b)_{crit}(0)- a ~\alpha^2$, and $a\approx 0.093$ is a numerical factor.

The defined  critical points  have been used to find if the amplitude of $\delta\sigma$ also scales as second power of $\alpha$. The expected scaling may be written as 
\begin{equation}
\frac{ \sigma(\alpha, T, M) - \sigma(0, T, M)}{\sigma(0, T, M)} \sim \alpha^n,
 \label{eq:scaling}
\end{equation}
where the expected power $n$ equals to 2. 
To check  this supposition we replotted the conductivity ratio from the Fig. \ref{fig:dsigma_ashift}  divided by $\alpha^2$  versus the shifted parameter $M/b -(M/b)_{crit}(\alpha)$.  
The result is depicted in Fig. \ref{fig:scaled_curves}. The scaling with power $n=2$ is only an approximate one.


\section{Conclusions}
\label{sec:7}

Our theory describes the holographic scenario of the quantum phase transition in $\zz$ Dirac semimetals, from the topologically non-trivial to the trivial phase. This system has been modelled by the action with Chern-Simons terms binding various 
combinations of $U(1)$-gauge field strengths with adequate coupling constants. The important ingredient, which we paid special attention to, is the direct coupling between both fields.
The coupling constant $\alpha$ between two gauge fields, corresponding to charge and spin degrees of freedom in Dirac semimetals with chiral anomaly and $\zz$ charge, 
has relatively small but well visible effect on the studied topological phase transition. Both the critical value of the $M/b$ and the magnitude of the Hall conductance do change with $\alpha$, but even 
if the changes are relatively small the effect is probably measurable.

Defining the dimensionless ratio of Hall conductivities, by the relation (\ref{deltasigma}), we observe that for low temperatures the mixing between both considered fields causes the strongly peaked shift in the studied Hall conductivity. On the contrary, fixing the temperature and varying the coupling strength parameter, one analyses the conductivity as a function of $M/b$. It can be noticed that the critical value of $M/b$ at which the phase transition happens
depends on the value of $\alpha$-coupling constant.

In the case of the longitudinal conductivity we have found that it is independent on parameters $M$ and $b$. Namely it has the constant value $\sigma_{xx} =\sigma_{yy} =\sigma_{zz} = r_0 =\pi T$. Such linear dependence on temperature is characteristic for the gap-less topological phase.


\acknowledgments
 The work of KIW has been supported by the National Science Centre (Poland) through the grant no. DEC-2017/27/B/ST3/01911.\\
We thank the unknown Referee for suggesting to analyze the conservation of the currents.

\appendix
\section{Constraints from the current conservation}
Now we pay some attention to the analysis of the boundary currents bounded with the Chern-Simons terms in the underlying theory and the conservation of currents.
In order to obtain some constraints on the coupling coefficients we shall study currents in our theory, i.e.,
one expands the holographic action about fixed background gauge fields of the form
$
A_\mu \rightarrow A_\mu + \delta A_\mu, ~  B_\mu \rightarrow B_\mu + \delta B_\mu,
$
to the second order in fluctuations \cite{gyn11}. After tedious calculations one finds that the variation of the action can be grouped into parts, one connected with the bulk action which is responsible for
the equations of motion and the boundary part from which we get expressions for the searched currents. They yield
\ben 
J^\alpha (F) &=& \frac{\delta S(gauge)}{\delta A_\alpha} \mid_{r \rightarrow \infty}
= \sqrt{-g} \Big( F^{\alpha r} + \frac{\alpha}{2} B^{\alpha r} \Big)\mid_{r \rightarrow \infty} \\ \nonumber
&+& \ep^{r \alpha \beta \ga \delta}
\Big( \frac{4}{3} \alpha_1  A_\beta F_{\ga \delta} + \frac{2}{3} \alpha_3  A_\beta B_{\ga \delta} + \frac{2}{3} \alpha_4  B_\beta B_{\ga \delta} \Big) \mid_{r \rightarrow \infty},\\ 
J^\alpha (B) &=& \frac{\delta S(gauge)}{\delta B_\alpha} \mid_{r \rightarrow \infty}
= \sqrt{-g} \Big( B^{ar} + \frac{\alpha}{2} F^{ar} \Big)\mid_{r \rightarrow \infty} \\ \nonumber
&+& \ep^{r \alpha \beta \ga \delta}
\Big( \frac{4}{3} \alpha_2 B_\beta B_{\ga \delta} + \frac{2}{3} \alpha_3  A_\beta F_{\ga \delta} + \frac{2}{3} \alpha_4 B_\beta F_{\ga \delta} \Big) \mid_{r \rightarrow \infty}.
\een

The important point is that in both currents we have mixture of gauge fields.  The current $J^\alpha (F)$ above corresponds to the charge current in the system and thus should be conserved. This immediately leads to $\alpha_1=\alpha_3=0$. On the other hand, the current $J^\alpha (B)$, which we interpret as the spin current does not vanish and is related to the $\zz$ anomaly of the studied model.
 
There is no obvious constraints on the couplings $\alpha_2$ and $\alpha_4$. In this context we add the following comment. It may be recalled that the appearance of the Chern-Simons terms, similar to those encountered in the action (\ref{action1}), has been discussed for the $(2+1)$-dimensional condensed matter system \cite{krive1992,sakhi1994}. The couplings have been found to depend on the lifetime $\tau$ of the fermions in the system. On the other hand, the lifetime changes with  the strength of disorder, electron-phonon interactions, etc. If a similar physics is realised in the considered $\zz$ Dirac systems the mentioned couplings may be system dependent.



\end{document}